\documentclass[11pt]{JHEP3}

\usepackage{amsmath,amssymb}
\usepackage{verbatim}
\usepackage{graphicx}
\usepackage{cite}

\DeclareFontFamily{OT1}{rsfs}{}
\DeclareFontShape{OT1}{rsfs}{m}{n}{ <-7> rsfs5 <7-10> rsfs7 <10->rsfs10}{} 
\DeclareMathAlphabet{\mycal}{OT1}{rsfs}{m}{n}

\renewcommand{\dag}{\dagger}

\newcommand{\be}[1]{ \begin{equation}\label{#1} }
\newcommand{\ee}{\end{equation}}
\newcommand{\bea}[1]{\begin{eqnarray}\label{#1} }
\newcommand{\eea}{\end{eqnarray}}

\newcommand{\eq}[2]{\begin{equation} #1 \label{#2} \end{equation}}

\newcommand{\de}{\delta}

\newcommand{\La}{\Lambda}

\newcommand{\mathsym}[1]{{}}

\newcommand{\cV}{\mathcal{V}}
\newcommand{\cW}{\mathcal{W}}

\newcommand{\pocha}[2]{\left(#1\right)_{#2}}
\newcommand{\pochd}[2]{\left[#1\right]_{#2}}

\newmuskip\pFqmuskip

\newcommand*\pFq[6][8]{%
  \begingroup 
  \pFqmuskip=#1mu\relax
  \mathcode`\,=\string"8000
  \begingroup\lccode`\~=`\,
  \lowercase{\endgroup\let~}\pFqcomma
  {}_{#2}F_{#3}{\left[\genfrac..{0pt}{}{#4}{#5};#6\right]}%
  \endgroup
}
\newcommand{\pFqcomma}{\mskip\pFqmuskip}

\title{Unitarity in three-dimensional flat space higher spin theories}

\author{D.~Grumiller, M.~Riegler and J.~Rosseel\\
           Institute for Theoretical Physics, 
           Vienna University of Technology,\\
           Wiedner Hauptstr. 8--10/136,
           A-1040 Vienna, Austria\\
           Email: \email{grumil, rieglerm, rosseelj@hep.itp.tuwien.ac.at}}

\abstract{
We investigate generic flat-space higher spin theories in three dimensions and find a no-go result, given certain assumptions that we spell out. Namely, it is only possible to have at most two out of the following three properties: unitarity, flat space,  non-trivial higher spin states. Interestingly, unitarity provides an (algebra-dependent) upper bound on the central charge, like $c=42$ for the Galilean $\mathcal{W}_4^{(2-1-1)}$ algebra.
We extend this no-go result to rule out unitary ``multi-graviton'' theories in flat space. We also provide an example circumventing the no-go result: Vasiliev-type flat space higher spin theory based on hs$(1)$ can be unitary and simultaneously allow for non-trivial higher-spin states in the dual field theory.
}

\keywords{flat space holography, higher spin theories, gravity in three dimensions, unitarity, Galilean conformal algebras, W-algebras}

\preprint{TUW--14--05}

\begin{document}











\section{Introduction}\label{se:1}

Interacting quantum field theories with massless fields of spin greater than $1/2$ are highly constrained by consistency requirements. This is in particular true for higher spin gauge theories, in which spins greater than $2$ are coupled to gravity and for which powerful no-go theorems (see for instance \cite{Bekaert:2010hw} for a review) rule out the existence of a non-trivial flat space S-matrix. The no-go theorems can be evaded by considering theories in curved backgrounds and indeed non-trivial higher-spin gauge theories in Anti-de~Sitter (AdS) or de~Sitter (dS) spacetimes \cite{Vasiliev:1990en} can be constructed (see \cite{Bekaert:2005vh,Didenko:2014dwa} for a review). Their flat space limit is however in general singular, in accord with the various no-go theorems.

Recently, an interacting spin 3 gauge theory in three flat spacetime dimensions was constructed \cite{Afshar:2013vka,Gonzalez:2013oaa}. This theory evades the above mentioned no-go theorems in a different way, namely by not possessing any propagating bulk degrees of freedom. This is made manifest by the fact that it can be formulated as a Chern-Simons theory, in complete analogy to the pure gravity case \cite{Achucarro:1987vz,Witten:1988hc}. The absence of bulk degrees of freedom does not necessarily imply that the theory is trivial; in particular in the presence of a boundary, boundary states that lie in a representation of a non-trivial asymptotic symmetry algebra can exist. As such, in  \cite{Afshar:2013vka,Gonzalez:2013oaa} it was shown that a choice of gauge group, along with consistent boundary conditions for the gauge field exist, such that the resulting Chern-Simons theory can be interpreted as a spin 3 gauge theory in flat space-time that exhibits a non-trivial, infinite-dimensional asymptotic symmetry algebra at future light-like infinity. The latter is an \.In\"on\"u--Wigner contraction of the ${\cal W}_3 \oplus {\cal W}_3$ algebra that was obtained by considering $\mathrm{SL}(3) \times \mathrm{SL}(3)$ Chern-Simons theory around AdS space-time \cite{Campoleoni:2010zq,Campoleoni:2011hg}. This is similar to how the $\mathrm{BMS}_3$ algebra, which appears as the asymptotic symmetry algebra for pure gravity around flat space, is obtained by contracting two copies of the Virasoro algebra. The non-linear algebra found in \cite{Afshar:2013vka,Gonzalez:2013oaa} can thus indeed be viewed as an appropriate spin 3 generalization of the $\mathrm{BMS}_3$ algebra. Although this example only corresponds to a spin 3 theory, it can be easily extended to more general three-dimensional flat space higher spin theories.

The work presented in \cite{Afshar:2013vka,Gonzalez:2013oaa} mainly focused on constructing the theories and resulting asymptotic symmetry algebras, without considering in detail whether or not they are compatible with unitarity. In the present work we investigate the issue of unitarity in 3-dimensional flat space higher spin gauge theories in more detail. With some notable exception, we find that generically only two out of the following three properties are possible: 1.~flat space, 2.~unitarity, 3.~non-trivial higher spin states. Thus, requiring unitarity in 3-dimensional flat space higher spin gauge theories is extremely restrictive. We will show this in higher spin theories in the principal embedding (such as the example of \cite{Afshar:2013vka,Gonzalez:2013oaa}), as well as in more general higher spin theories, in non-principal embeddings. We will moreover argue that this absence of unitarity in non-trivial flat space higher spin theories is a rather generic consequence of the structure of central charges and non-linearities in their algebras. In fact, up to simple extensions the only unitary example with non-trivial higher spin states we were able to find is a specific Vasiliev-type higher spin gauge theory, that is a Chern-Simons theory with gauge algebra hs$(1)$. This example evades the above mentioned restrictions precisely because the resulting flat space asymptotic symmetry algebra can be linearized. Indeed, this symmetry algebra can be viewed as an \.In\"on\"u--Wigner contraction of two copies of the linear ${\cal W}_\infty$ algebra \cite{Pope:1989ew,Pope:1989sr}.

This paper is organized as follows. In section \ref{se:2} we recapitulate flat space/Galilean contractions of conformal algebras (Virasoro and ${\cal W}$-algebras) and restrictions on the central charges from unitarity.
In section \ref{se:3} we contract two copies of the Polyakov--Bershadsky (quantum) algebra and derive restrictions from unitarity on the central charge.
In section \ref{se:4} we prove a general no-go result for flat space contractions of higher spin theories.
In section \ref{se:5} we generalize the discussion to Vasiliev-type of higher spin theories and provide a counterexample to our no-go result by circumventing one of its premises.
In section \ref{se:6} we conclude.

\section{Flat space contractions and unitarity}\label{se:2}

In this section we summarize known facts about flat space contractions and unitarity. In subsection \ref{se:2.1} we review aspects of 3-dimensional flat space holography for spin-2 theories, following Bagchi et al., and Barnich et al. In subsection \ref{se:2.2} we comment on unitarity for Galilean conformal algebras. In subsection \ref{se:2.3} we collect some of the main results for the \.In\"on\"u--Wigner contraction of higher spin gravity for the principal embedding \cite{Afshar:2013vka,Gonzalez:2013oaa}.

\subsection{Flat space holography}\label{se:2.1}

There are two ways to set up flat space holography. Either one just formulates everything in flat space --- the boundary conditions, field equations, classical solutions, boundary charges, asymptotic symmetry algebra, etc.~--- or one starts from (A)dS and takes the limit of vanishing cosmological constant, $\La=\pm1/\ell^2\to 0$. The fact that dS has rather different properties from AdS is a caveat that sometimes the limit can be subtle. However, at the level of symmetry algebras it is straightforward --- and useful --- to consider specifically AdS in the limit of infinite AdS radius.
\eq{
\textrm{Flat\;Space:}\quad \ell\to\infty
}{eq:fshsg1}
Let us now start with the asymptotic symmetry algebra of (quantum) gravity in AdS$_3$. Following Brown and Henneaux, this symmetry algebra is the conformal algebra in two dimensions \cite{Brown:1986nw}:
\begin{align}
[{\cal L}_n,\,{\cal L}_m] &= (n-m)\,{\cal L}_{n+m}+\frac{c}{12}\,(n^3-n)\,\de_{n+m,\,0}\\ 
[\bar {\cal L}_n,\,\bar {\cal L}_m] &= (n-m)\,\bar {\cal L}_{n+m}+\frac{\bar c}{12}\,(n^3-n)\,\de_{n+m,\,0}
\label{eq:fshsg2}
\end{align}
Following Barnich and Comp\`ere \cite{Barnich:2006av} as well as Bagchi, Gopakumar and collaborators \cite{Bagchi:2009my,Bagchi:2009pe}, we then combine the Virasoro generators linearly in either of the following two ways ($\epsilon=1/\ell\ll 1$).
\begin{align}
& \textrm{Galilean\;contraction:} \; && L_n := {\cal L}_n + \bar{\cal L}_n\quad && M_n :=  -\epsilon \,\big({\cal L}_n - \bar{\cal L}_n\big) \label{eq:fshsg3} \\
& \textrm{Ultrarelativistic\;contraction:} \; && L_n := {\cal L}_n - \bar{\cal L}_{-n}\quad && M_n :=  \epsilon \,\big({\cal L}_n + \bar{\cal L}_{-n}\big) \label{eq:fshsg4} 
\end{align}
In either of these cases the contracted algebra (after taking the limit $\epsilon\to 0$) is isomorphic to the BMS$_3$ algebra \cite{Barnich:2006av},\footnote{
See the recent papers by Duval, Gibbons and Horvathy for a higher-dimensional discussion \cite{Duval:2014uoa,Duval:2014uva,Duval:2014lpa}.
} which in turn is isomorphic to the Galilean conformal algebra \cite{Bagchi:2010zz} and contains the three-dimensional Poincar\'{e} algebra as global subalgebra:
\begin{subequations}
 \label{eq:fshsg5}
\begin{align}
 [L_n,\,L_m]&=(n-m)\,L_{n+m}+\frac{c_L}{12}\,(n^3-n)\,\de_{n+m,\,0}\\ 
 [L_n,\,M_m]&=(n-m)\,M_{n+m}+\frac{c_M}{12}\,(n^3-n)\,\de_{n+m,\,0}\\
 [M_n,\,M_m]&=0
\end{align}
\end{subequations}
The central charges $c_{L,\,M}$ depend on the type of contraction:
\begin{align}
& \textrm{Galilean\;contraction:} \; && c_L = c+\bar c \quad && c_M = \lim_{\epsilon\to 0} \epsilon (-c+\bar c) \label{eq:fshsg6} \\
& \textrm{Ultrarelativistic\;contraction:} \; && c_L = c-\bar c \quad && c_M = \lim_{\epsilon\to 0} \epsilon (c+\bar c) \label{eq:fshsg7} 
\end{align}
Note in particular that the $c_M$ central charge is dimensionful, as it is proportional to the contraction parameter $\epsilon$. 

Physically, the correct contraction from AdS to flat space is the ultrarelativistic one (see \cite{Bagchi:2012cy} for a discussion): in terms of Carter--Penrose diagrams, the asymptotic AdS cylinder gets boosted to null infinity. Algebraically, however, the Galilean contraction is somewhat simpler, since there is no mixing between operators of positive and negative conformal weights in the definitions \eqref{eq:fshsg3}. Since we are mostly interested in asymptotic symmetry algebras and their most general central extensions we shall therefore employ the Galilean contractions whenever they are simpler to perform.

While the main focus of our paper is unitarity in higher spin theories, we mention here briefly some of the recent achievements in flat space holography, many of which should be generalizable to higher spin theories. First steps towards a generalization to four dimensions were pursued by Barnich and Troessaert in \cite{Barnich:2010eb,Barnich:2011mi}. A direct flat space limit of Brown--Henneaux boundary conditions and the spectrum of physical (as well as some unphysical) states with zero mode charges switched on was provided by Barnich, Gomberoff and Gonzalez \cite{Barnich:2012aw}. This includes in particular (shifted-boost) orbifolds of flat space, so-called flat space cosmologies \cite{Cornalba:2002fi,Simon:2002ma,Cornalba:2003kd}. A first concrete suggestion for a holographic dual to a specific flat space gravity theory (dubbed `flat space chiral gravity') was given in \cite{Bagchi:2012yk}, where conformal Chern--Simons gravity at level $k=1$ was conjectured to be dual to a chiral half of the monster CFT (with $c=24$), in the spirit of Witten as well as Li, Song and Strominger \cite{Witten:2007kt,Li:2008dq}. A microscopic counting of the states responsible for the entropy of flat space cosmology solutions was provided independently by Barnich \cite{Barnich:2012xq} and by Bagchi, Detournay, Fareghbal and Simon \cite{Bagchi:2012xr}. The flat limit of Liouville theory, the theory controlling the classical boundary dynamics in AdS$_3$, was taken by Barnich, Gomberoff and Gonzalez \cite{Barnich:2012rz} (see also \cite{Barnich:2013yka}). The existence of a phase transition between (hot, rotating) flat space and flat space cosmologies, as well as consistency with the chiral Cardy formula for flat space chiral gravity was shown in \cite{Bagchi:2013lma}. Afshar provided flat space boundary conditions of conformal gravity in the Chern--Simons formulation \cite{Afshar:2013bla}. Schulgin and Troost realized the BMS$_3$ algebra in terms of vertex operators on the string worldsheet \cite{Schulgin:2013xya}. A discussion of holographic currents in flat space holography was provided by Strominger \cite{Strominger:2013lka,He:2014laa} and Barnich and Troessaert \cite{Barnich:2013axa}, with one of the key insights being that one can use currents and their algebras even in the absence of conserved charges (see also \cite{Banks:2014iha,Barnich:2013sxa,Wald:1999wa}). Aspects of flat space holography in the presence of a non-backreacting scalar field were studied by Costa \cite{Costa:2013vza} and independently by Fareghbal and Naseh \cite{Fareghbal:2013ifa}. Logarithmic corrections to the Galilean Cardy and Bekenstein--Hawking formulas were calculated by Bagchi and Basu \cite{Bagchi:2013qva}. 
A reinterpretation of flat space \.In\"on\"u--Wigner contractions like in \eqref{eq:fshsg3}, \eqref{eq:fshsg4} was provided by Krishnan, Raju and Roy \cite{Krishnan:2013wta} by replacing the parameter $\epsilon$ formally by a Grassmanian number, which automatically excludes terms quadratic in $\epsilon$ and was applied to discuss desingularization of the Milne Universe within spin-3 gravity \cite{Krishnan:2013tza}. Most recently, a well-defined variational principle for flat space Einstein gravity was provided and exploited to calculate 0- and 1-point functions in \cite{Detournay:2014fva}.

\subsection{Unitarity in Galilean conformal algebras}\label{se:2.2}

We focus now on the Galilean conformal algebra \eqref{eq:fshsg5} and check under which conditions it allows unitary representations. In order to proceed we have to choose a vacuum. Our choice is defined by the highest weight conditions
\eq{
L_n|0\rangle = M_n|0\rangle = 0 \quad \forall n \geq -1\,.
}{eq:fshsg8}
While the vacuum conditions \eqref{eq:fshsg8} seem pretty standard and lead to a Poincar\'e-invariant vacuum, it is nevertheless debatable if they are always the right choice. This is particularly true for the ultrarelativistic contraction \eqref{eq:fshsg4}, where the generators $L_n$, $M_n$ mix positive and negative weight generators of the original CFT. We shall not enter such a debate here and always stick to the vacuum definition \eqref{eq:fshsg8} (or appropriate higher spin generalizations thereof) in the present work. With similar caveats, we define also hermitian conjugation in a standard way,
\eq{
L_n^\dagger := L_{-n}\qquad M_n^\dagger := M_{-n}\,.
}{eq:fshsg17}

Having defined the vacuum and hermitian conjugation, it is straightforward to address the issue of unitarity. Constructing the inner products of all level-2 descendants yields the Gram matrix
\eq{
\left(\begin{array}{ll}
\langle0|L_2 L_{-2}|0\rangle = \frac{c_L}{2} & \quad \langle0|L_2 M_{-2}|0\rangle = \frac{c_M}{2}\\
\langle0|M_2 L_{-2}|0\rangle = \frac{c_M}{2} & \quad \langle0|M_2 M_{-2}|0\rangle = 0
\end{array}\right)
}{eq:fshsg9}
whose determinant is $-\tfrac{c_M^2}{4}$. Thus, if $c_M\neq 0$ then there is always a positive and a negative norm state, regardless of the signs of the central charges. Therefore, as long as $c_M\neq 0$ the Galilean conformal algebra \eqref{eq:fshsg5} does not have any unitary representations [provided we stick to the vacuum \eqref{eq:fshsg8} and hermitian conjugation \eqref{eq:fshsg17}; this caveat applies everywhere in our paper]. Note that this argument also applies to more general algebras that contain the BMS$_3$ algebra as a subalgebra.

For $c_M=0$ the Gram matrix \eqref{eq:fshsg9} has vanishing determinant, which means there is at least one null state. Assuming $c_L\neq 0$ (otherwise the algebra becomes trivial) there is exactly one null state and one state whose norm depends on the sign of $c_L$. In this case one can truncate the whole sector of $M_{-n}$ descendants of the vacuum, since they are all null states \cite{Bagchi:2009pe,Bagchi:2012yk}, and one is left with just a single copy of the Virasoro algebra and the corresponding Virasoro descendants of the vacuum, $L_{-n}|0\rangle$. Then standard CFT considerations of unitary representations of the Virasoro algebra apply; in particular, the central charge $c_L$ must be positive for unitarity. 

In conclusion, necessary conditions for unitarity of Galilean conformal algebras \eqref{eq:fshsg5} with vacuum \eqref{eq:fshsg8} and hermitian conjugation \eqref{eq:fshsg17} are
\eq{
c_M=0\qquad c_L\geq 0\,.
}{eq:fshsg10}
Requiring non-triviality converts the inequality into a strict one, $c_L>0$.

\subsection{Contraction for higher spin theories in principal embedding ($\mathcal{W}_N$)}\label{se:2.3}

The contractions \eqref{eq:fshsg3}, \eqref{eq:fshsg4} can be generalized to ${\cal W}$-algebras, which appear as asymptotic symmetry algebras in higher spin theories in AdS$_3$ \cite{Henneaux:2010xg,Campoleoni:2010zq}. These theories are most efficiently formulated as Chern--Simons gauge theories, typically with gauge group SL$(N)\times$SL$(N)$, with specific boundary conditions imposed on the gauge connection. Which ${\cal W}$-algebra one obtains depends, among other things, on the way the SL$(2)$ part describing gravity is embedded into the SL$(N)$. In this section we focus on the principal embedding, in which the theory has higher spin fields with spins $2,\, 3,\,\dots,\, N$. The ensuing asymptotic symmetry algebra is denoted by ${\cal W}_N$. Since we want to keep our discussion theory-independent at a purely algebraic level, we are not going to write down any actions, specific theories or explicit results for the central charges. We just assume there is some theory that leads to the algebras we present. This assumption is justified in all cases that we are going to discuss, at least for specific sets of choices for the central charges.

The simplest case is spin-3 AdS gravity, whose asymptotic symmetry algebra consists of two copies of the ${\cal W}_3$ algebra \cite{Henneaux:2010xg,Campoleoni:2010zq}. The non-trivial commutation relations between the generators of a single copy of ${\cal W}_3$ are given by \cite{Zamolodchikov:1985wn,Mathieu:1988pm,Bakas:1989mx}
\begin{align}
 [{\cal L}_n,\,{\cal L}_m] &= (n-m)\,{\cal L}_{n+m}+\frac{c}{12}\,(n^3-n)\,\de_{n+m,\,0} \\
 [{\cal L}_n,\,{\cal W}_m] &= (2n-m)\,{\cal W}_{n+m} \\
 [{\cal W}_n,\,{\cal W}_m] &= (n-m)(2n^2+2m^2-nm-8)\,{\cal L}_{n+m} + \frac{96}{c+\tfrac{22}{5}}\,(n-m)\,\colon\!{\cal L}{\cal L}\colon_{n+m} \nonumber \\ 
&\quad + \frac{c}{12}\,(n^2-4)(n^3-n)\,\de_{n+m,\,0}
\end{align}
with the usual normal ordering prescription
\eq{
\colon\!{\cal L}{\cal L}\colon_{n} = \sum_{p\geq -1} {\cal L}_{n-p}{\cal L}_p + \sum_{p< -1} {\cal L}_p{\cal L}_{n-p} - \frac{3}{10} (n+3)(n+2) {\cal L}_n\,.
}{eq:fshsg11}
The generators of the other copy of ${\cal W}_3$ will be denoted with bar on top, $\bar{\cal L}_n$ and $\bar{\cal W}_n$.

We define the Galilean contraction by analogy to the spin-2 case \eqref{eq:fshsg3}:
\begin{subequations}
\label{eq:fshsg12}
		\begin{align}
			L_n:=&\,\mathcal{L}_n+\bar{\mathcal{L}}_{n}\qquad & M_n &:=-\epsilon\left(\mathcal{L}_n-\bar{\mathcal{L}}_{n}\right)\\
			U_n:=&\,{\cal W}_n+\bar{{\cal W}}_{n}\qquad & V_n &:=-\epsilon\left({\cal W}_n-\bar{{\cal W}}_{n}\right)
		\end{align}
\end{subequations}
The contracted algebra was derived in \cite{Afshar:2013vka}:
\begin{subequations}
 \label{eq:FSHSG10}
\begin{align}
[L_n,\, L_m] &= (n-m) L_{n+m} + \frac{c_L}{12}\, (n^3 - n)\, \delta_{n+m,\,0}  \\
[L_n,\, M_m] &= (n-m) M_{n+m} + \frac{c_M}{12}\, (n^3 - n)\, \delta_{n+m,\,0}  \\
[L_n,\, U_m] &= (2n-m) U_{n+m}  \\
[L_n,\, V_m] &= (2n-m) V_{n+m}  \\
[M_n,\, U_m] &= (2n-m) V_{n+m}  \displaybreak[1] \\
[U_n,\, U_m] &= (n-m)(2 n^2 + 2 m^2 - nm -8) L_{n+m} 
+\frac{192}{c_M}\,  (n-m) \Lambda_{n+m}  \\ & \quad - \frac{96\big(c_L + \tfrac{44}{5}\big)}{c_M^2}\,  (n-m) \Theta_{n+m} 
+\frac{c_L}{12}\, n(n^2-1)(n^2-4)\, \delta_{n+m,\,0} \label{eq:FSHSG20} \\
[U_n,\, V_m] &= (n-m)(2 n^2 + 2 m^2 - nm -8) M_{n+m} 
+ \frac{96}{c_M}\,  (n-m) \Theta_{n+m} \nonumber \\
& \quad +\frac{c_M}{12}\, n(n^2-1)(n^2-4)\, \delta_{n+m,\,0} 
\end{align}
We used the definitions 
\eq{
\Theta_n  =\sum_p M_p\, M_{n-p}\qquad
\Lambda_n =\sum_p\colon\! L_p\, M_{n-p}\colon -\tfrac{3}{10}(n+2)(n+3)M_n
}{eq:fshsg18}
and normal ordering
\eq{
\colon\! L_n\, M_m\colon = L_n\, M_m \textrm{\;if\;} n<-1 \qquad \colon\! L_n\, M_m\colon = M_m\, L_n \textrm{\;if\;}n\geq -1\,.
}{eq:fshsg13}
\end{subequations}
The central charges are given by \eqref{eq:fshsg6} in terms of the original central charges $c$ and $\bar c$. We assume here that $c_L$ and $c_M$ can take arbitrary (real) values.

We address now unitarity. The vacuum is defined again by the conditions \eqref{eq:fshsg8}, supplemented by
\eq{
U_n|0\rangle=V_n|0\rangle=0\qquad\forall n \geq -2\,.
}{eq:fshsg14}
Similarly, the hermitian generators \eqref{eq:fshsg17} are supplemented by $U_n^\dagger:=U_{-n}$, $V_n^\dagger:=V_{-n}$.
The spin-2 result for level-2 descendants \eqref{eq:fshsg9} still applies, so that unitarity again requires the necessary conditions \eqref{eq:fshsg10}. 

As discussed in \cite{Afshar:2013vka}, the condition $c_M=0$ leads to a further contraction of the algebra \eqref{eq:FSHSG10}. This is so due to the appearance of first and second order poles in $c_M$ in the commutation relations of flat space higher spin generators $U_n$ and $V_n$. Singularities are avoided if we rescale the generators
\eq{
U_n \to c_M U_n
}{eq:fshsg15}
before taking the limit $c_M\to 0$. The contracted algebra then simplifies and the non-vanishing commutators read
\begin{subequations}
 \label{eq:fshsg16}
\begin{align}
[L_n,\, L_m] &= (n-m) L_{n+m} + \frac{c_L}{12}\, (n^3 - n)\, \delta_{n+m,\,0}  \\
[L_n,\, M_m] &= (n-m) M_{n+m} \\
[L_n,\, U_m] &= (2n-m) U_{n+m}  \\
[L_n,\, V_m] &= (2n-m) V_{n+m}  \\
[U_n,\, U_m] &\propto [U_n, V_m] = 96(n-m)\,\Theta_{n+m}
\end{align}
\end{subequations}
As in the spin-2 case, the remaining non-trivial part of the algebra is a single copy of the Virasoro algebra (if $c_L\neq 0$). In particular, all the descendants of higher spin generators $U_{-n}$, $V_{-n}$ (and of the supertranslations $M_{-n}$) are null states. This means that at least for the present example unitarity in flat space higher spin gravity leads to an elimination of all physical higher spin states. The only physical states that arise as descendants of the vacuum are the usual Virasoro descendants.

An interesting question is to what extent our main conclusion depends on the specific ${\cal W}$-algebra that we used. We address this question first for the principal embedding.

Precisely the same conclusions are reached for arbitrary ${\cal W}_N$ algebras. A first step in this direction is to generalize the contraction \eqref{eq:fshsg14} to all higher spin generators, which is straightforward \cite{Gonzalez:2013oaa}. It can be checked that the contracted algebra again contains poles in $c_M$, which requires an additional contraction like in \eqref{eq:fshsg15}. The final flat space algebra compatible with unitarity is essentially of the form \eqref{eq:fshsg16}: again, all higher spin states decouple and the only physical states that arise as descendants of the vacuum are the usual Virasoro descendants.

In the remainder of the paper we generalize the discussion to non-principal embeddings and also to the hs$(\lambda)$ case.

\section{Unitarity in contracted Polyakov--Bershadsky ($\mathcal{W}_3^{(2)}$)}\label{se:3}

It is interesting to extend the discussion of the previous section to more general higher spin theories, i.e, to drop the assumption that we are in the principal embedding. In this section we focus on the simplest non-principal embedding, which leads to the Polyakov--Bershadsky algebra, $\mathcal{W}^{(2)}_3$, as asymptotic symmetry algebra.

In section \ref{se:3.1} we contract two copies of the Polyakov--Bershadsky (quantum) algebra to its Galilean (quantum) version.
In section \ref{se:3.2} we discuss restrictions on the central charge from unitarity and find non-trivial upper and lower bounds.

\subsection{Galilean Polyakov--Bershadsky}\label{se:3.1}
 
The (quantum) $\mathcal{W}_3$ algebra is generated by $\mathcal{L}_n$, $\hat{{\cal G}}^{\pm}_n$ and ${\cal J}_n$ whose non-vanishing commutation relations read \cite{Polyakov:1989dm,Bershadsky:1990bg}
\begin{subequations}\label{eq:W32AlgebraQuantumBershadsky}
\begin{align}
[\mathcal{L}_n,\,\mathcal{L}_m]=&\,(n-m)\mathcal{L}_{n+m}+\frac{c}{12}\,n(n^2-1)\,\delta_{n+m,\,0}\\
[\mathcal{L}_n,\,{\cal J}_m]=&\,-m{\cal J}_{n+m}\\
[\mathcal{L}_n,\,\hat{{\cal G}}_m^{\pm}]=&\,\big(\frac{n}{2}-m\big)\hat{{\cal G}}_{n+m}^{\pm}\\
[{\cal J}_n,\, {\cal J}_m]=&\, \frac{2k+3}{3}\,n\,\delta_{n+m,\,0}\\
[{\cal J}_n,\,\hat{{\cal G}}_m^{\pm}]=&\,\pm\hat{{\cal G}}_{n+m}^{\pm}\\
[\hat{{\cal G}}_n^{+},\,\hat{{\cal G}}_m^{-}]=&\,-(k+3)\mathcal{L}_{n+m}+\frac{3}{2}(k+1)(n-m){\cal J}_{n+m}
+3:\!{\cal JJ}\!:_{\,n+m}\nonumber\\
&+\frac{(k+1)(2k+3)}{2}\,\big(n^2-\frac{1}{4}\big)\,\delta_{n+m,\,0}
\end{align}
\end{subequations}
with the central charge
\begin{equation}\label{eq:CentralCharge}
c=25-\frac{24}{k+3}-6(k+3) 
\end{equation}
and the normal ordering prescription
\begin{equation}
:\!{\cal JJ}\!:_{\,n}=\sum_{p\geq0}{\cal J}_{n-p}{\cal J}_p+\sum_{p<0}{\cal J}_p{\cal J}_{n-p}\, .
\end{equation}
By replacing $\mathcal{L}_n\rightarrow\bar{\mathcal{L}}_n$, $\hat{{\cal G}}^{\pm}_n\rightarrow\hat{\bar{{\cal G}}}^{\pm}_n$, ${\cal J}_n\rightarrow\bar{{\cal J}}_n$, $k\rightarrow\bar{k}$ and $c\rightarrow\bar{c}$ in \eqref{eq:W32AlgebraQuantumBershadsky} one obtains the commutation relations for the second copy of the $\mathcal{W}^{(2)}_3$ algebra needed for the Galilean contraction. In order to properly contract these two algebras a rescaling of $\hat{{\cal G}}_n$ and $\hat{\bar{{\cal G}}}_n$ with a suitable factor, e.g.~$\sqrt{-k-1}$, is necessary. Otherwise terms of $\mathcal{O}(\frac{1}{\epsilon})$ would spoil the limit $\epsilon\rightarrow0$. We drop the hat for the rescaled generators, $\hat {\cal G}^\pm_n = \sqrt{-k-1}\,{\cal G}^\pm_n$ and similarly for $\bar{\mathcal G}_n^\pm$.

The linear combinations that will lead to the Galilean contraction of the Polyakov--Bershadsky algebra are defined analog to \eqref{eq:fshsg3} and \eqref{eq:fshsg12}
\begin{subequations}
\begin{align}\label{eq:NRLinComb}
L_n:=&\,\mathcal{L}_n+\bar{\mathcal{L}}_{n}\qquad & M_n &:=-\epsilon\left(\mathcal{L}_n-\bar{\mathcal{L}}_{n}\right)\\
J_n:=&\,{\cal J}_n+\bar{{\cal J}}_{n}\qquad & K_n &:=-\epsilon\left({\cal J}_n-\bar{{\cal J}}_{n}\right)\\
U^{\pm}_n:=&\,{\cal G}^{\pm}_n+\bar{{\cal G}}^{\pm}_{n}\qquad & V^{\pm}_n &:=-\epsilon\left({\cal G}^{\pm}_n-\bar{{\cal G}}^{\pm}_{n}\right)\,.
\end{align}
\end{subequations}
Actually, there are some ambiguities in the normalizations of the generators $G^\pm$, $\bar G^\pm$, $U^\pm$ and $V^\pm$, which we fix in convenient ways and with no loss of generality.

The limit $\epsilon\rightarrow0$ then yields the contracted algebra
\begin{subequations}\label{eq:FlatW32Quantum}
\begin{align}
[L_n,\,L_m]=&\,(n-m)L_{n+m}+\frac{c_L}{12}\,n(n^2-1)\,\delta_{n+m,\,0}\\
[L_n,\,M_m]=&\,(n-m)M_{n+m}+\frac{c_M}{12}\,n(n^2-1)\,\delta_{n+m,0} \displaybreak[1] \\
[L_n,\,J_m]=&\,-mJ_{n+m}\\
[L_n,\,K_m]=&\,-mK_{n+m}  \displaybreak[1] \\
[L_n,\,U^{\pm}_m]=&\,\big(\frac{n}{2}-m\big)\,U^\pm_{n+m}\\
[L_n,\,V^{\pm}_m]=&\,\big(\frac{n}{2}-m\big)\,V^\pm_{n+m} \displaybreak[1]  \\
[M_n,\,J_m]=&\,-mK_{n+m}\\
[M_n,\,U^\pm_m]=&\,\big(\frac{n}{2}-m\big)\,V^\pm_{n+m}  \displaybreak[1] \\
[J_n,\,J_m]=&\,\frac{32-c_L}{9}\,n\,\delta_{n+m,\,0}\label{eq:FlatW32QuantumO1}\\
[J_n,\,K_m]=&\,-\frac{c_M}{9}\,n\,\delta_{n+m,\,0}  \displaybreak[1] \\	
[J_n,\,U^\pm_m]=&\,\pm  U^\pm_{n+m}\\
[J_n,\,V^\pm_m]=&\,\pm V^\pm_{n+m}\\
[K_n,\,U^\pm_m]=&\,\pm V^\pm_{n+m}  \displaybreak[1] \\
[U^+_n,\,U^-_m]=&\,L_{n+m}-\frac{3}{2}(n-m)J_{n+m}-\frac{18\left(c_L-26\right)}{c_M^2}:\!KK\!:_{\,n+m}+\frac{18}{c_M}:\!JK\!:_{\,n+m}\nonumber\\
&+\frac{\left(c_L-32\right)}{6}\,\big(n^2-\frac{1}{4}\big)\,\delta_{n+m,\,0}\\
[U^\pm_n,\,V^\mp_m]=&\,\pm M_{n+m}-\frac{3}{2}(n-m)K_{n+m}\pm
\frac{18}{c_M}\,:\!KK\!:_{\,n+m}\pm \frac{c_M}{6}\,\big(n^2-\frac{1}{4}\big)\,\delta_{n+m,\,0}		
\end{align}
\end{subequations}
with the central charges
\eq{
c_L = c+\bar{c}\qquad c_M=\epsilon(\bar{c}-c)
}{eq:angelinajolie}
and the normal ordering prescription
\begin{align}
		&:\!KK\!:_{\,n}\,\equiv\,\sum_{p\geq0}K_{n-p}K_p+\sum_{p<0}K_pK_{n-p}\\
		&:\!JK\!:_{\,n}\,\equiv\,\sum_{p\geq0}\left(K_{n-p}J_p+J_{n-p}K_p\right)+\sum_{p<0}\left(J_pK_{n-p}+K_pJ_{n-p}\right)\,.
\end{align}
We call the algebra \eqref{eq:FlatW32Quantum} `Galilean Polyakov--Bershadsky' algebra and will use a similar nomenclature for other Galilean contractions of two copies of ${\cal W}$-algebras discussed in later sections.

\subsection{Unitarity in Galilean Polyakov--Bershadsky}\label{se:3.2}

For the same reason as in the spin-2 case there are no unitary representations of the algebra \eqref{eq:FlatW32Quantum} for $c_M\neq0$. Thus, if we want unitary representations we have to take the limit $c_M\rightarrow0$, which requires that we rescale the generators appropriately.
	\begin{equation}
		U^\pm_n\rightarrow\hat{U}^\pm_n=c_MU^\pm_n\qquad
		V^\pm_n\rightarrow\hat{V}^\pm_n=c_MV^\pm_n
	\end{equation}
Taking the limit $c_M\rightarrow0$ leads to a further contraction of the Galilean Polyakov--Bershadsky algebra.
The non-vanishing commutators are given by
	\begin{subequations}\label{eq:CM0Algebra}
		\begin{align}
			[L_n,L_m]=&\, (n-m)L_{n+m}+\frac{c_L}{12}\,n(n^2-1)\delta_{n+m,\,0} \label{eq:gpb1}\\
			[L_n,J_m]=&\, -mJ_{n+m}\\
			[J_n,J_m]=&\, \frac{32-c_L}{9}\,n\,\delta_{n+m,\,0} \label{eq:gpb2}\\
			[L_n,M_m]=&\, (n-m)M_{n+m}\\
			[L_n,K_m]=&\, [M_n,J_m]=-mK_{n+m} \displaybreak[1] \\
			[L_n,\hat{U}^{\pm}_m]=&\, \big(\frac{n}{2}-m\big)\,\hat{U}^\pm_{n+m}\\
			[L_n,\hat{V}^{\pm}_m]=&\, [M_n,\hat{U}^\pm_m]=\big(\frac{n}{2}-m\big)\,\hat{V}^\pm_{n+m} \displaybreak[1] \\
			[J_n,\hat{U}^\pm_m]=&\, \pm \hat{U}^\pm_{n+m}\\
			[J_n,\hat{V}^\pm_m]=&\, [K_n,\hat{U}^\pm_m]=\pm\hat{V}^\pm_{n+m}\\
			[\hat{U}^+_n,\hat{U}^-_m]=&\, -18\left(c_L-26\right)\,:\!KK\!:_{\,n+m}\,.
		\end{align}
	\end{subequations}

To discuss unitarity we define the hermitian conjugates of our operators 
	\begin{gather}
		L_n^\dagger:=L_{-n}\qquad M_n^\dagger:=M_{-n}\qquad J_n^\dagger:=J_{-n}\qquad K_n^\dagger:=K_{-n}\\
		\left(U^\pm_n\right)^\dagger:=U^\mp_{-n}\qquad\left(V^\pm_n\right)^\dagger:=V^\mp_{-n}
	\end{gather}
and the vacuum in the usual way
	\begin{align}
		L_n|0\rangle=&M_n|0\rangle=0\qquad\{\forall n\in\mathbb{Z}|n\geq-1\}\\
		J_m|0\rangle=&K_ m|0\rangle=0\qquad\{\forall m\in\mathbb{Z}|m\geq0\}\\
		U^\pm_p|0\rangle=&V^\pm_p|0\rangle=0\qquad\{\forall p\in\mathbb{Z}+\frac{1}{2}|p\geq-\frac{1}{2}\}\,.
	\end{align}

We discuss now restrictions on the remaining central charge $c_L$ from demanding unitarity.
The central terms in the Virasoro algebra \eqref{eq:gpb1} and the current algebra \eqref{eq:gpb2} must have non-negative signs. This immediately implies lower and upper bounds on the central charge.
\eq{
0 \leq c_L \leq 32
}{eq:lalapetz}

States generated by $M_n$, $K_n$, $U^\pm_n$ and $V^\pm_n$ have zero norm and are orthogonal to all other states. Thus we can mod out these states and extend our definition of the vacuum in the following way
	\begin{align}
		L_n|0\rangle=&J_{n+1}|0\rangle=0\qquad\{\forall n\in\mathbb{Z}|n\geq-1\}\\
		M_m|0\rangle=&K_ m|0\rangle=0\qquad\forall m\in\mathbb{Z}\\
		U^\pm_p|0\rangle=&V^\pm_p|0\rangle=0\qquad\forall p\in\mathbb{Z}+\frac{1}{2}\,.
	\end{align}
The only (perturbative) states that remain in the theory for $0\leq c_L\leq32$ are descendants of the vacuum $L_{-n}|0\rangle$ with $n>1$, $J_{-m}|0\rangle$ with $m>0$ or combinations thereof. In order to have a well defined basis of states at level $N$, we employ the following ordering of operators
	\begin{equation}
		J_{-n_1}^{m_1}\ldots J_{-n_{p}}^{m_p}L_{-n_{p+1}}^{m_{p+1}}\ldots L_{-n_N}^{m_N}|0\rangle\,,
	\end{equation}
with 
		$m_i\in\mathbb{N}$, 
		$n_1,\ldots,n_p\in\mathbb{N}\backslash\{0\}$,
		$n_{p+1},\ldots,n_N\in\mathbb{N}\backslash\{0,1\}$,
		$n_1>\ldots>n_p$,
		$n_{p+1}>\ldots>n_N$, and
		$\sum_{i=1}^Nm_in_i=N$.
We check now further restrictions from non-negativity of the norm of descendants of the vacuum.

At level 1 there is only one state generated by $J_{-1}$. The norm of this state is given by $C_J=-\frac{(c_L-32)}{9}$ which is non-negative if the bound on the central charge \eqref{eq:lalapetz} holds.

At level 2 three states generated by $L_{-2}$, $J_{-2}$ and $J_{-1}^2$ are present. The Gram matrix $K^{(2)}$ is given by
	\begin{equation}
		K^{(2)}=\left(
			\begin{array}{lll}
				  \langle0|L_2L_{-2}0\rangle=\frac{c_L}{2} & \quad \langle0|L_2J_{-2}|0\rangle =0 & \quad \langle0|L_2J_{-1}^2|0\rangle=C_J \\
				\langle0|J_2L_{-2}|0\rangle = 0 & \quad \langle0|J_2J_{-2}|0\rangle=2C_J & \quad \langle0|J_2J_{-1}^2|0\rangle=0\\
				\langle0|J_1^2L_{-2}|0\rangle=C_J & \quad \langle0|J_1^2J_{-2}|0\rangle=0& \quad \langle0|J_1^2J_{-1}^2|0\rangle=2C_J^2
			\end{array}\right)\,.
\label{eq:gpb4}
	\end{equation}
For $C_J=0$ the two $J$-descendants are null states. For $c_L=0$ descendants generated by $L_n$ have zero norm but are not orthogonal to all other states in the theory, so they are not null states and cannot be modded out. 
Two of the three eigenvalues of the Gramian \eqref{eq:gpb4}, $\lambda_0=2C_J$ and $\lambda_+=\tfrac{1}{4}\big(c_L+4 C_J^2+\sqrt{(c_L+4 C_J^2)^2+(1-c_L)16 C_J^2}\big)$ are non-negative in the whole range $0\leq c_L\leq32$. The third eigenvalue, $\lambda_-=\tfrac{1}{4}\big(c_L+4 C_J^2 - \sqrt{(c_L+4 C_J^2)^2+(1-c_L)16 C_J^2}\big)$, changes its sign at $c_L=1$ and is positive in the range $1\leq c_L\leq32$. At $c_L=1$ the descendant associated to $L_{-2}$ is proportional to the $J_{-1}^2$ descendant.
	\begin{equation}
		L_{-2}|0\rangle=\tfrac{9}{62}\,J_{-1}^2|0\rangle
	\end{equation} 
We have checked explicitly that the key features discussed above persist for level 3 and 4 descendants of the vacuum. In particular, non-negativity of the eigenvalues of the Gramian always restricts to values of $c_L$ larger or equal to $1$. The same is expected to hold for arbitrary levels.

Thus, the region $0\leq c_L< 1$ is excluded and necessary conditions deduced from the analysis above for the central charge to be consistent with unitarity are
\eq{
1\leq c_L \leq 32\,.
}{eq:gpb3}
At the lower end of the allowed interval, $c_L=1$, the determinant of the Gram matrix vanishes and thus some states become linearly dependent. Only $J_{-n}$ descendants remain in the theory and all $L_{-n}$ descendants depend linearly on them. It is noteworthy that this is also the value where the Polyakov--Bershadsky algebra has its only non-trivial unitary representation \cite{Afshar:2012nk}. At the upper end of the allowed interval, $c_L=32$, the states corresponding to the $\hat{\mathfrak{u}}(1)$ part of the algebra become null states and only the Virasoro modes remain. The resulting representation is unitary for our choice of the vacuum.

While there are further restrictions from unitarity, our results above show already two remarkable features: 
\begin{enumerate}
\item Requiring unitarity implies that all higher-spin descendants of the vacuum become null states and drop out of the physical spectrum. 
\item Unitarity implies a lower and upper bound on the central charge, in this case $1\leq c_L\leq 32$.
\end{enumerate}
We consider now further restrictions from unitarity. To this end we make the inverse Sugawara-shift 
\eq{
T = L - \frac{1}{2k}\,\colon\! JJ\!\colon
}{eq:new1}
with the $\hat{\mathfrak{u}}(1)$ level $k=\tfrac{9}{c_L-32}$ from \eqref{eq:CM0Algebra}, so that the CFT factorizes into two commuting CFTs, a free boson with central charge $c=1$ and a coset Virasoro CFT with central charge $c=c_L-1$. Unitarity of the latter restricts $c$ to the minimal model values $c=0,\,\tfrac12,\,\tfrac{7}{10},\,\dots$ for $c<1$ \cite{diFrancesco}.
In the next section we generalize the results of this section to generic flat space higher spin theories.

\section{No-go: Unitarity in contractions of general non-linear $\mathcal{W}$-algebras}\label{se:4}

We have seen in section \ref{se:2} that flat space higher spin theory cannot be unitary for the principal embedding, unless we contract to a theory where all higher spin states become null states. Moreover, we have generalized these results to the Galilean Polyakov--Bershadsky algebra in section \ref{se:3}. The way in which the contraction works makes it plausible that also other non-principal embeddings could lead to similar conclusions. In this section we show that this is indeed the case, first by considering particular examples and then by proving a general no-go result.

In section \ref{se:4.2} we discuss a particular spin-4 example that leads to an upper bound $c_L=42$ for the central charge.
In section \ref{se:4.3} we give an infinite family of higher spin examples in the next-to-principal embedding.
In section \ref{se:4.1} we prove a general no-go result.
In section \ref{se:4.4} we show that our no-go result also allows to eliminate multi-graviton states in flat space.

\subsection{Upper bound on central charge ($\mathcal{W}_4^{(2-1-1)}$)}\label{se:4.2}

In this section we show that the existence for an upper bound on the central charge arises not only in the Polyakov--Bershadsky algebra, but also in the Galilean contraction derived from the $\mathcal{W}_4^{(2-1-1)}$ algebra. We start with two copies of this algebra, generated by $\mathcal{L}_n,J_n,S^\pm_n,S^0_n,G^{a|b}_n$, with $a,b=\pm$,
	\begin{subequations}
		\begin{align}
			[\mathcal{L}_n,\,\mathcal{L}_m]=&\,(n-m)\mathcal{L}_{n+m}+\frac{c}{12}\,n(n^2-1)\,\delta_{n+m,\,0}\\
			[\mathcal{L}_n,\,S^{a}_m]=&\,-mS^{a}_{n+m}\\
			[\mathcal{L}_n,\,J_m]=&\,-mJ_{n+m}\\
			[\mathcal{L}_n,\,G^{a|b}_m]=&\,\big(\frac{n}{2}-m\big)\,G^{a|b}_{n+m} \displaybreak[1] \\
			[S^{a}_n,\,S^{b}_m]=&\,(a-b)S_{n+m}^{a+b}-\frac{k+1}{2}\,\big(1-3a^2\big)\,n\,\delta_{a+b,\,0}\,\delta_{n+m,\,0}\label{QA:NastyComm}\\
			[S^{a}_n,\,G^{b|d}_m]=&\,\frac{a-b}{2}\,G^{2a+b|d}_{n+m} \displaybreak[1] \\
			[J_n,\,G^{a|b}_m]=&\,2bG^{a|b}_{n+m}\\
			[J_n,\,J_m]=&-4kn\,\delta_{n+m,\,0} \displaybreak[1] \\
			[G^{\pm|\pm}_n,\,G^{\pm|\mp}_m]=&\,\frac{k}{\alpha k+\beta}\,(n-m)S^{\pm}_{n+m}\mp\frac{1}{\alpha k+\beta}\,:\!JS\!:^{\,\pm}_{\,n+m} \displaybreak[0] \\
			[G^{a|\pm}_n,\,G^{-a|\mp}_m]=&\,a\frac{(k-2)}{\alpha k+\beta}\,\mathcal{L}_{n+m}\mp\frac{a}{2}\,\frac{(k+1)}{\alpha k+\beta}\,(n-m)J_{n+m}+\frac{k}{\alpha k+\beta}\,(n-m)S_{n+m}^{0}\nonumber\\
			&a\frac{k(k+1)}{\alpha k+\beta}\,\big(n^2-\frac{1}{4}\big)\,\delta_{n+m,\,0}-\frac{a}{\alpha k+\beta}\,\big(:\!SS\!:^{\,\{-|+\}}_{\,n+m}\nonumber\\
			&-\frac{3}{8}\,:\!JJ\!:_{\,n+m}\mp b\,:\!JS\!:^{\,0}_{\,n+m}-2\,:\!SS^{0|0}\!:_{\,n+m}\big)		
		\end{align}	
	\end{subequations}
with the central charge
	\begin{equation}
		c=\frac{3(k+2)(2k+1)}{k-2}\,,
	\end{equation}
and  $\alpha,\beta\in\mathbb{R}$.
We define the following linear combinations
	\begin{subequations}
		\begin{align}
			L_n:=\,&\mathcal{L}_n+\bar{\mathcal{L}}_{n}\quad & M_n &:=-\epsilon\left(\mathcal{L}_n-\bar{\mathcal{L}}_{n}\right)\\
			O_n:=&\,J_n+\bar{J}_{n}\quad  &P_n &:=-\epsilon\left(J_n-\bar{J}_{n}\right)\\
			Q^{a}_n:=&\,S^{a}_n+\bar{S}^{a}_{n}\quad & R^{a}_n &:=-\epsilon\big(S^{a}_n-\bar{S}^{a}_{n}\big)\\
			U^{a|b}_n:=&\,G^{a|b}_n+\bar{G}^{a|b}_{n}\quad & V^{a|b}_n &:=-\epsilon\big(G^{a|b}_n-\bar{G}^{a|b}_{n}\big)\,.
		\end{align}
	\end{subequations}
In the limit $\epsilon\rightarrow0$ we obtain the following non-vanishing linear commutation relations
	\begin{subequations}\label{eq:Flat211}
		\begin{align}
			[L_n,\,L_m]=&\,(n-m)L_{n+m}+\frac{c_L}{12}\,n(n^2-1)\,\delta_{n+m,\,0}\\
			[L_n,\,M_m]=&\,(n-m)M_{n+m}+\frac{c_M}{12}\,n(n^2-1)\,\delta_{n+m,\,0}\\
			[L_n,\,O_m]=&\,-mO_{n+m}\\
			[L_n,\,P_m]=&\,-mP_{n+m} \displaybreak[1] \\
			[L_n,\,Q^{a}_m]=&\,-mQ^{a}_{n+m}\\
			[L_n,\,R^{a}_m]=&\,-mR^{a}_{n+m} \displaybreak[1] \\
			[L_n,\,U^{a|b}_m]=&\,\big(\frac{n}{2}-m\big)U^{a|b}_{n+m}\\
			[L_n,\,V^{a|b}_m]=&\,\big(\frac{n}{2}-m\big)V^{a|b}_{n+m} \displaybreak[1] \\
			[M_n,\,O_m]=&\,-mP_{n+m}\\
			[M_n,\,Q^{a}_m]=&\,-mR^{a}_{n+m}\\
			[M_n,\,U^{a|b}_m]=&\,\big(\frac{n}{2}-m\big)V^{a|b}_{n+m} \displaybreak[1] \\
			[O_n,\,O_m]=&\,\frac{2(54-c_L)}{3}\,n\,\delta_{n+m,\,0}\\
			[O_n,\,P_m]=&\,-\frac{2c_M}{3}\,n\,\delta_{n+m,\,0} \displaybreak[1] \\	
			[O_n,\,U^{a|b}_m]=&\,2bU^{a|b}_{n+m}\\
			[O_n,\,V^{a|b}_m]=&\,2bV^{a|b}_{n+m}\\
			[P_n,\,U^{a|b}_m]=&\,2bV^{a|b}_{n+m} \displaybreak[1] \\
			[Q^{a}_n,\,Q^{b}_m]=&\,(a-b)Q^{a+b}_{n+m}+\frac{42-c_L}{12}\,(1-3a^2)n\,\delta_{a+b,\,0}\,\delta_{n+m,\,0}\label{eq:Flat211OaOb}\\
			[Q^{a}_n,\,R^{b}_m]=&\,(a-b)R^{a+b}_{n+m}-\frac{c_M}{12}\,(1-3a^2)n\,\delta_{a+b,\,0}\,\delta_{n+m,\,0} \displaybreak[1] \\
			[Q^{a}_n,\,U^{b|d}_m]=&\,\frac{a-b}{2}U^{2a+b|d}_{n+m}\label{eq:Flat211OaUpmb}\\
			[Q^{a}_n,\,V^{b|d}_m]=&\,\frac{a-b}{2}V^{2a+b|d}_{n+m}\\
			[R^{a}_n,\,U^{b|d}_m]=&\,\frac{a-b}{2}V^{2a+b|d}_{n+m}
		\end{align}
	\end{subequations}
and the following non-linear relations
	\begin{subequations}
		\begin{align}
			[U^{\pm|\pm}_n,\,U^{\pm|\mp}_m]=&\,(n-m)\big(\frac{1}{\alpha}Q^{\pm}_{n+m}-\frac{12\beta}{\alpha^2c_M}R^{\pm}_{n+m}\big)\mp\frac{6}{\alpha c_M}\big(:\!OR\!:^{\,\pm}_{\,n+m}+:\!PQ\!:^{\,\pm}_{\,n+m}\big)\nonumber\\
			&\pm\frac{6(\alpha c_L+12\beta-54\alpha)}{\alpha^2c_M^2}\,:\!PR\!:^{\,\pm}_{\,n+m} \displaybreak[1] \\
			[U^{a|\pm}_n,\,U^{-a|\mp}_m]=&\,a\big(\frac{1}{\alpha}L_{n+m}-\frac{12(2\alpha+\beta)}{\alpha^2c_M}M_{n+m}\big)\mp\frac{a}{2}(n-m)\big(\frac{1}{\alpha}O_{n+m}+\frac{12(\alpha-\beta)}{\alpha^2c_M}P_{n+m}\big)\nonumber\\
			&+(n-m)\big(\frac{1}{\alpha}Q^{0}_{n+m}-\frac{12\beta}{\alpha^2c_M}R^{0}_{n+m}\big)+a\frac{\alpha c_L-6(7\alpha+2\beta)}{6\alpha^2}\,\big(n^2-\frac{1}{4}\big)\,\delta_{n+m,\,0}\nonumber\\
			&-\frac{6}{c_M}\,\big(\{:\!QR\!:\}^{\,\{+|-\}}_{\,n+m}-\tfrac{3}{8}\{:\!OP\!:\}_{\,n+m}\pm a(:\!OR\!:^{\,0}_{\,n+m}+:\!PQ\!:^{\,0}_{\,n+m})\nonumber\\
			&-2\{:\!QR\!:\}^{\,0|0}_{\,n+m}\big)+\frac{6(ac_L+12b-54a)}{ac_M^2}\,\big(2\,:\!RR\!:^{\,\{-|+\}}_{\,n+m}\nonumber\\
			&-\frac{3}{8}\,:\!PP\!:_{\,n+m}\pm a\,:\!PR\!:^{\,0}_{\,n+m}-2\,:\!RR\!:^{\,0|0}_{\,n+m}\big) \displaybreak[1] \\
			[U^{\pm|\pm}_n,\,V^{\pm|\mp}_m]=&\,(n-m)\frac{1}{\alpha}R^{\pm}_{n+m}\mp\frac{6}{\alpha c_M}\,:\!PR^{\pm}\!:_{\,n+m}\nonumber\\
			[U^{a|\pm]}_n,\,V^{-a|\mp}_m]=&\,\frac{a}{\alpha}M_{n+m}\mp\frac{a}{2\alpha}(n-m)P_{n+m}+(n-m)\frac{1}{\alpha}R^{0}_{n+m}+a\frac{c_M}{6\alpha}\big(n^2-\frac{1}{4}\big)\,\delta_{n+m,\,0}\nonumber\\
			&-\frac{6a}{\alpha c_M}\,\big(2\,:\!RR\!:^{\,\{+|-\}}_{\,n+m}-\frac{3}{8}\,:\!PP\!:_{\,n+m}\pm a\,:\!PR\!:^{\,0}_{\,n+m}-2\,:\!RR\!:^{\,0|0}_{\,n+m}\big)\,.						
		\end{align}
	\end{subequations}
As in the Galilean Polyakov--Bershadsky case we have to require $c_M=0$ and the ``colored states have to be rescaled with $c_M$. 

Besides the BMS algebra the contracted algebra contains an affine $\hat{\mathfrak{su}}(2)$ \eqref{eq:Flat211OaOb} subalgebra, so one has to be careful how to properly define hermitean conjugation of the corresponding operators. If one simply assumed the following prescription for hermitean conjugation of the operators $Q^a_n$ (and the standard one for the other operators)
	\begin{equation}
	\textrm{wrong\;hermitean\;conjugation:}	\left(Q^a_n\right)^\dag=Q^a_{-n}\,,
	\end{equation}
then by looking at the norm of the first level descendants one would conclude that there is only one possible value of $c_L$ where unitary representations are possible. This is so, because the central terms of the $[Q^0_n,Q^0_m]$ and $[Q^\pm_n,Q^\mp_m]$ commutators have different signs. One can, however, solve this problem by a redefinition of the hermitean conjugate of the following operators as ($\hat{U}^{\pm|a}_n=c_MU^{\pm|a}_n$ and $\hat{V}^{\pm|a}_n=c_MV^{\pm|a}_n$)
	\begin{align}
		\left(Q^0_n\right)^\dag=&Q^0_{-n}\qquad\left(Q^\pm_n\right)^\dagger=\gamma Q^\mp_{-n}\\
		\left(R^0_n\right)^\dag=&R^0_{-n}\qquad\left(R^\pm_n\right)^\dagger=\gamma R^\mp_{-n}\\
		\left(\hat{U}^{\pm|a}_n\right)^\dag=&\mu^\pm\nu^a\hat{U}^{\mp|-a}_{-n}\qquad\left(\hat{V}^{\pm|a}_n\right)^\dag=\mu^\pm\nu^a \hat{V}^{\mp|-a}_{-n}\,,
	\end{align} 
where $\gamma,\mu^\pm$ and $\nu^\pm$ are some real numbers which will be determined by demanding consistency with the contracted algebra at hand. In order to do that we first look at
	\begin{equation}
		\left([Q^\pm_n,\,Q^\mp_m]\right)^\dagger 
                =-\gamma^2[Q^\mp_{-n},\,Q^\pm_{-m}]
		=\pm2Q^0_{-(n+m)}-\frac{42-c_L}{6}\, n\,\delta_{n+m,\,0}\,,
	\end{equation}
which tells us that that $\gamma^2=1$ in order to satisfy \eqref{eq:Flat211OaOb}. We choose $\gamma=-1$ so that the norm of the states $Q^0_{-n}|0\rangle$ and $Q^\pm_{-n}|0\rangle$ have the same sign. We check now whether or not a choice for $\mu^\pm$ and $\nu^\pm$ exists that is compatible with our algebra. 
	\begin{equation}
		\big(\big[Q^\pm_n,\,\hat{U}^{\mp|a}_m\big]\big)^\dagger 
                =\mu^\mp\nu^a\big[Q^\mp_{-n},\,\hat{U}^{\pm|-a}_{-m}\big]
		=\pm\big(\hat{U}^{\pm|a}_{n+m}\big)^\dagger=\pm\mu^\pm\nu^a\hat{U}^{\mp|-a}_{-(n+m)}
	\end{equation}
In order to be consistent with \eqref{eq:Flat211OaUpmb} the parameters $\mu^\pm$ must satisfy
	\begin{equation}
		\frac{\mu^+}{\mu^-}=-1\,.	
	\end{equation}
The last thing to be checked is the rescaled $[\hat{U}^{\pm|\pm}_n,\,\hat{U}^{\pm|\mp}_m]$ commutator. Proceeding in the same manner as before we find
	\begin{equation}
		\nu^+\nu^-=-1\,.
	\end{equation}
Thus we can consistently define hermitean conjugation in such a way that states generated by operators associated with the affine $\hat{\mathfrak{su}}(2)$ subalgebra do not necessarily have to be excluded.

Looking at the Gramian matrix of the first levels descendants of the vacuum gives us more constraints on possible values of $c_L$. The representation can be unitary in the interval
	\begin{equation}
		29-\sqrt{661}\leq c_L\leq 42\,.
\label{eq:whatever}
	\end{equation}
The lower bound may look funny, but has a simple explanation if we write the central charge as sum of three terms:
\eq{
c_L = c_{\hat{\mathfrak{u}}(1)} + c_{\hat{\mathfrak{su}}(2)} + c_{\textrm{\tiny bare}}
}{eq:crest}
The first term on the right hand side comes from the $\hat{\mathfrak{u}}(1)$ current algebra and gives a contribution $c_{\hat{\mathfrak{u}}(1)}=1$. Using the standard expression $c=k \,\textrm{dim}\, g/(k+h^\vee)$, where in our case dim$\,g=3$, level $k=(42-c_L)/6$ and dual Coxeter number $h^\vee=2$, yields for the second term $c_{\hat{\mathfrak{su}}(2)}=3(42-c_L)/(54-c_L)$. Thus, we can express the `bare' central charge as
\eq{
c_{\textrm{\tiny bare}} = c_L - 1 - \frac{3(42-c_L)}{54-c_L}\,.
}{eq:crest1}
Requiring non-negativity of the bare central charge, $c_{\textrm{\tiny bare}}\geq 0$, translates into the condition $c_L\geq 29-\sqrt{661}\approx3.290$, which nicely explains the lower bound in \eqref{eq:whatever}. 

Looking more closely at the norm of the first few descendants we find a gap between the upper limit, $c_L=42$, and the next highest value of $c_L$ compatible with unitarity, $c_L=36$. In terms of the $\hat{\mathfrak{su}}(2)$ level this corresponds to the gap between $k=0$ and $k=1$. There will be further conditions from unitarity\footnote{%
For instance, restrictions on unitarity from the $\hat{\mathfrak{su}}(2)_k$ algebra part imply that $k$ has to be a positive integer, which --- together with the inequalities \eqref{eq:whatever} --- means that only the discrete set $c_L=6n$ with $n=1,\,2,\,\dots,\,7$ is allowed.
} 
that restrict the allowed values of $c_L$ in the interval \eqref{eq:whatever}, but it is interesting that already at this stage we get both lower and upper bounds on the central charge, similar to section \ref{se:3.2}.


\subsection{Feigin--Semikhatov family of examples ($\mathcal{W}_N^{(2)}$)}\label{se:4.3}

We construct now the Galilean contraction of two copies of the Feigin--Semikhatov algebra, which provides an infinite family of examples. Since we are interested in unitary representations of the resulting algebras we will only take a detailed look at the resulting central charges. Starting with two copies of the following $\mathcal{W}^{(2)}_N$ algebra
	\begin{subequations}
		\begin{align}
			[J_n,\,J_n]=&\kappa n\delta_{n+m,\,0}\\
			[J_n,\,\mathcal{L}_m]=&nJ_{n+m}\\
			[J_n,\,G^\pm_m]=&\pm G^\pm_{n+m} \displaybreak[0] \\
			[\mathcal{L}_n,\,\mathcal{L}_m]=&(n-m)\mathcal{L}_{n+m}+\frac{c}{12}n(n^2-1)\delta_{n+m,\,0}\\
			[\mathcal{L}_n,\,G_m^{\pm}]=&\big(n(\tfrac{N}{2}-1)-m\big)\,G_{n+m}^{\pm} \displaybreak[0] \\
			[G_n^{+},\,G_m^{-}]=&\frac{\lambda_{N-1}(N,k)}{(N-1)!} \,f(n)\,\delta_{n+m,\,0}+g(n,m)\lambda_{N-2}(N,k)J_{n+m}\ldots\\
			[\hat{W}^s_n,\,\textnormal{anything}]=&\ldots
		\end{align}
	\end{subequations} 
with 
	\begin{align}
		&\kappa=\frac{N-1}{N}k+N-2\\
		&c=-\frac{((N+k)(N-1)-N)((N+k)(N-2)N-N^2+1)}{N+k}\\
		&\lambda_n(N,k)=\prod_{i=1}^n(i(k+N-1)-1)
	\end{align}
and $f(n)$ and $g(n,m)$ being some functions of their respective arguments whose explicit form does not matter for the following discussion. In order to have a well defined contraction in the limit $\epsilon\rightarrow0$ we first have to rescale some of the generators in an appropriate way. The main hurdle for having a well defined contraction is the $k$-behavior of some of the structure constants. If we parametrize the central charges of the two copies in the following way
	\begin{equation}
		c_i=\frac{1}{2\epsilon}\left(\alpha c_M+\beta\epsilon c_L\right)
	\end{equation}
with $\alpha=\beta=1$ for $c_1\equiv c$ and $\alpha=-\beta=-1$ for $c_2\equiv\bar{c}$ then $k_1\equiv k$ ($k_2\equiv\bar{k}$) is for $\epsilon\rightarrow0$ approximated by the following expression
	\begin{equation}
		k_i\sim-\frac{\alpha c_M}{2N(2-3N+N^2)\epsilon}-\frac{\beta  c_L+2 (N (N ((N-5) N+5)+1)-1)}{2N(N-2) (N-1) }+\mathcal{O}(\epsilon)
	\end{equation}
which means that every power of $k$ ($\bar{k}$) is proportional to $\frac{c_M}{\epsilon}$. This in turn tells us that the central terms can at most be polynomials of degree one in terms of $k$ ($\bar{k}$) and all other structure constants can at most be of degree zero in $k$ ($\bar{k}$). Thus this rescaling only applies to the higher spin generators $\hat{W}^s_n$ and the generators $G^\pm_n$. After rescaling we can define the following linear combinations
	\begin{subequations}
		\begin{align}\label{eq:NRLinCombFS}
			L_n:=&\mathcal{L}_n+\bar{\mathcal{L}}_{n} &&\quad M_n:=-\epsilon\big(\mathcal{L}_n-\bar{\mathcal{L}}_{n}\big)\\
			O_n:=&J_n+\bar{J}_{n}&&\quad K_n:=-\epsilon\big(J_n-\bar{J}_{n}\big)\\
			U^{\pm}:=&G^{\pm}_n+\bar{G}^{\pm}_{n}&&\quad V^{\pm}:=-\epsilon\big(G^{\pm}_n-\bar{G}^{\pm}_{n}\big)\\
			W^s_n:=&\hat{W}^s_n+\bar{\hat{W}}^s_{n}&&\quad X^s_n:=-\epsilon\big(\hat{W}^s_n-\bar{\hat{W}}^s_{n}\big)
		\end{align}
	\end{subequations}
and calculate the resulting algebra in the limit $\epsilon\rightarrow0$. This leads to the following algebra
	\begin{subequations}\label{eq:FlatFSQuantum}
		\begin{align}
			[L_n,L_m]=&(n-m)L_{n+m}+\frac{c_L}{12}n(n^2-1)\delta_{n+m,0}\\
			[L_n,M_m]=&(n-m)M_{n+m}+\frac{c_M}{12}n(n^2-1)\delta_{n+m,0} \displaybreak[1] \\
			[L_n,O_m]=&-mO_{n+m}\\
			[L_n,K_m]=&-mK_{n+m} \displaybreak[1] \\
			[L_n,U^{\pm}_m]=&\left(\frac{n}{2}-m\right)U^\pm_{n+m}\\
			[L_n,V^{\pm}_m]=&\left(\frac{n}{2}-m\right)V^\pm_{n+m} \displaybreak[1] \\
			[M_n,O_m]=&-mK_{n+m}\\
			[M_n,U^\pm_m]=&\left(\frac{n}{2}-m\right)V^\pm_{n+m} \displaybreak[1] \\
			[O_n,O_m]=&-\frac{2 (N-1)^2 (N+1)- c_L}{(N-2) N^2}n\delta_{n+m,0}\\
			[O_n,K_m]=&-\frac{c_M}{(N-2) N^2}n\delta_{n+m,0} \displaybreak[1] \\	
			[O_n,U^\pm_m]=&\pm  U^\pm_{n+m}\\
			[O_n,V^\pm_m]=&\pm V^\pm_{n+m}\\
			[K_n,U^\pm_m]=&\pm V^\pm_{n+m} \displaybreak[1] \\
			[U^+_n,U^-_m]=&\ldots\label{eq:FlatFSQuantumUU}\\
			[U^\pm_n,V^\mp_m]=&\ldots\label{eq:FlatFSQuantumUV}\\
			[W^s_n,\textnormal{anything}]=&\ldots\label{eq:FlatFSQuantumWAll}\\
			[X^s_n,\textnormal{anything}]=&\ldots\label{eq:FlatFSQuantumXAll}		
		\end{align}
	\end{subequations}
with the central charges \eqref{eq:fshsg6}.

Since we are mainly interested in unitary representations of these algebras we have to look again at the $c_M\rightarrow0$ limit. By construction all central terms in \eqref{eq:FlatFSQuantumUU}-\eqref{eq:FlatFSQuantumXAll} are either proportional to $c_L$ or $c_M$. In addition the commutators in \eqref{eq:FlatFSQuantumUU}-\eqref{eq:FlatFSQuantumXAll} have non-linear terms proportional to powers of $\frac{1}{c_M}$ which again makes it necessary to rescale the generators $U^\pm_n$, $V^\pm_n$, $W^\ell_n$, $X^\ell_n$ by appropriate powers of $c_M$. After doing this the limit $c_M\to 0$ eliminates all the central terms in \eqref{eq:FlatFSQuantumUU}-\eqref{eq:FlatFSQuantumXAll}. The only issue we still have to worry about is whether or not terms proportional to powers of $L_{n+m}$ an $O_{n+m}$ can appear on the right hand side of \eqref{eq:FlatFSQuantumUU}-\eqref{eq:FlatFSQuantumXAll}. If that were the case then we could not simply mod out the states since the inner product of theses states with other states would yield a non-zero result and thus we could not dispose of them. There is however a simple argument as to why such terms cannot appear. Half of the generators defined via the contraction acquire an additional dimension of inverse length via the parameter $\epsilon=\frac{1}{\ell}$ with $\ell$ being the AdS radius. Thus terms proportional to powers purely consisting of $L_{n+m}$ and $O_{n+m}$ cannot carry inverse powers of $c_M$, which also is dimensionful. The only possibility for the appearance of terms that are powers of $L_{n+m}$ and $O_{n+m}$ and carry inverse powers of $c_M$ is via mixing with $M_{n+m}$, $P_{n+m}$, $V^\pm_{n+m}$ and $X^s_{n+m}$. Such terms only appear as cross terms during the contraction and as such are not the one with the highest inverse power of $c_M$ on the right hand side of \eqref{eq:FlatFSQuantumUU}-\eqref{eq:FlatFSQuantumXAll}. Thus they are gone after a rescaling with the highest power of $c_M$ appearing on the right hand side of the commutators in question. This in turn means that all the higher-spin states and descendants $G_{-n}^\pm|0\rangle$ are null states for unitary representations of the Galilean Feigin--Semikhatov algebras. 

Repeating the same analysis as in the Polyakov--Bershadsky case we see that demanding the absence of negative norm states the range of possible values of  $c_L$ that could allow for unitary representations is given by
	\begin{equation}
		1\leq c_L\leq2(N-1)^2(N+1)\,.
	\end{equation}
This means that the maximal value of $c_L$ for which we obtain a unitary CFT grows as $N^3$ for large $N$ in contrast to linear growth in $N$ in the uncontracted Feigin--Semikhatov case. Note that all these CFTs are chiral in the sense that their symmetry algebra contains a single copy of the Virasoro algebra. 

\subsection{General no-go result}\label{se:4.1}

The previous examples all point towards the conclusion that, under the assumptions we have been working with, it is not possible to find unitary representations of flat space higher spin algebras that contain non-trivial higher-spin states. In this section, we will argue, on dimensional grounds, that this conclusion is generic for flat space higher spin algebras that are inherently non-linear and can be obtained via \.In\"on\"u--Wigner contractions of AdS higher spin $\mathcal{W}$-algebras.

Suppose therefore that we start from two copies of an inherently non-linear $\mathcal{W}$-algebra. Such an algebra contains higher spin generators that we will denote by $W_n$ ($\bar{W}_n$ for the second copy), whose commutation relations can schematically be written as:
\begin{subequations}
 \label{eq:WWComm}
	\begin{align}
		[W_n,\,W_m]&=\ldots+f(c)\,:\!AB\!:_{\,n+m}+\ldots+\omega(c)\!\!\!\!\!\!\prod_{j=-(s-1)}^{s-1}\!\!\!\!\!(n+j)\,\delta_{n+m,\,0}\,, \\
		[\bar{W}_n,\,\bar{W}_m]&=\ldots+f(\bar{c})\,:\!\bar{A}\bar{B}\!:_{\,n+m}+\ldots+\omega(\bar{c})\!\!\!\!\!\!\prod_{j=-(s-1)}^{s-1}\!\!\!\!\!(n+j)\,\delta_{n+m,\,0}\,.
	\end{align}
\end{subequations}
These commutators contain non-linear terms that involve other generators of the algebra, denoted by $A_n$, $B_n$ ($\bar{A}_n$, $\bar{B}_n$ for the second copy). Note that for some cases, the $A_n$ generators can be the same as the $B_n$ generators, but this does not necessarily have to be the case. The two copies of the $\mathcal{W}$-algebra are characterized by central charges $c$, $\bar{c}$. These appear in the non-linear terms and in the central charge terms in \eqref{eq:WWComm} via functions $f(c)$ and $\omega(c)$. We will keep these functions arbitrary for the sake of generality, apart from the following restriction on $f(c)$ [and similarly for $f(\bar c)$]:
\begin{equation} \label{restrf}
\lim_{c \rightarrow \infty} f(c)  = 0 \,,
\end{equation}
i.e., requiring that the non-linear terms are not important in the semi-classical regime of large central charges. 
This property holds for all examples that were constructed explicitly in the literature so far, and one can plausibly argue that it holds generically.
The ellipses in \eqref{eq:WWComm} denote additional linear and other non-linear terms which can possibly appear on the right-hand side of the commutator.  

Starting from these two copies one can obtain a flat space higher spin algebra via a Galilean contraction, involving a contraction parameter $\epsilon$ of dimension [length]$^{-1}$. We thus define the new generators as before:
\begin{align}
U_n:=&\,W_n+\bar{W}_n\,,\quad &V_n &:=-\epsilon\left(W_n-\bar{W}_n\right)\,,\\
C_n:=&\,A_n+\bar{A}_n\,,\quad &D_n &:=-\epsilon\left(A_n-\bar{A}_n\right)\,,\\
E_n:=&\,B_n+\bar{B}_n\,,\quad &F_n &:=-\epsilon\left(B_n-\bar{B}_n\right) \,.
\end{align} 
We also define again the central charges as
\begin{equation}
c_L = c + \bar{c} \,, \qquad \qquad c_M = \epsilon( \bar{c} - c) \,.
\end{equation}
From these redefinitions, one can already infer that the generators $U_n$, $C_n$, $E_n$, as well as the central charge $c_L$ are dimensionless. Similarly, the generators $V_n$, $D_n$, $F_n$ and the central charge $c_M$ have dimensions [length]$^{-1}$.

In order to obtain unitarity under the assumptions we are working with, we have to take the limit $c_M \rightarrow 0$ in the flat space higher spin algebra that results from the above contraction, for the reasons explained in section \ref{se:2.2}. The crux of the no-go argument lies in showing that this limit can not be taken in the commutators of the flat space higher spin generators $U_n$, $V_n$, in such a way that non-trivial central terms remain. The higher spin generators $U_n$, $V_n$ can therefore only create null states, which can be modded out.

For the commutator $[U_n, V_m]$, this is immediate on dimensional grounds. This commutator has dimensions of [length]$^{-1}$ and in order to have the same dimension, the central charge term necessarily has to be proportional to $c_M$, implying that the $[U_n, V_m]$ commutator will be center-less in the limit $c_M\rightarrow0$. The commutator $[V_n, V_m]$ is zero upon contraction, so the only non-trivial commutator to examine is $[U_n, U_m]$.
Let us first look at the structure of the non-linear terms. Performing the contraction, one obtains
	\begin{align}\label{UULimit}
		\lim_{\epsilon\rightarrow0}[U_n,\,U_m]=&\,\ldots+\lim_{\epsilon\rightarrow0}\frac{1}{4\epsilon^2}\,\big((f(c)+f(\bar{c}))(\epsilon^2\,:\!CE\!:_{\,n+m}+\,:\!DF\!:_{\,n+m})\nonumber\\
		&+\frac{(f(\bar{c})-f(c))}{\epsilon}(:\!CF\!:_{\,n+m}+\,:\!DE\!:_{\,n+m})\big)+\ldots
	\end{align}
For a generic function $f(c)$, such that \eqref{restrf} holds, one finds that
	\begin{equation}
		f(c)+f(\bar{c})\sim\mathcal{O}(\epsilon^2)\quad\textnormal{and}\quad f(c)-f(\bar{c})\sim\mathcal{O}(\epsilon) \,.
	\end{equation}
The contraction $\epsilon\rightarrow0$ can thus consistently be performed and one sees that the non-linear terms that generically survive the contraction are the ones of the form $:\!DF\!:_{\,n+m}$, $:\!CF\!:_{\,n+m}$ and $:\!DE\!:_{\,n+m}$. These have dimensions [length]$^{-2}$, [length]$^{-1}$ and [length]$^{-1}$, respectively. Since the commutator $[U_n, U_m]$ is dimensionless, these non-linear terms have to appear with prefactors that depend on $c_M$, in order to compensate for their dimensions. The structure of the $[U_n,U_m]$ commutator is thus schematically given by 
\begin{align}\label{eq:UUContracted}
		[U_n,\,U_m]&=\ldots+\mathcal{O}(\frac{1}{c_M^2})\,:\!DF\!:_{\,n+m}+\mathcal{O}(\frac{1}{c_M})\left(:\!CF\!:_{\,n+m}+\,:\!DE\!:_{\,n+m}\right) \nonumber \\ &\qquad +\tilde{\omega}(c_L) \!\!\!\!\!\!\prod_{j=-(s-1)}^{s-1}\!\!\!\!\!(n+j)\,\delta_{n+m,\,0},
\end{align}
where the central term has to be a function of $c_L$ on dimensional grounds.

In order to take the limit $c_M \rightarrow 0$, we thus have to rescale $U_n\rightarrow\tilde{U}_n:=c_MU_n$ in the case bilinear terms are present (and no terms of higher order than bilinear terms). The central terms in the $[\tilde{U}_n, \tilde{U}_m]$ commutator however do not survive this rescaling, showing that all higher spin generators of the flat space algebra lead to null states when acting on the vacuum. 

Although we have given the argument for the case in which the non-linear terms are bilinear, it can easily be extended to the case where non-linear terms of higher order appear, by rescaling the generators  $U_n$ in accordance with the highest negative power of $c_M$ appearing on the right hand side of \eqref{eq:UUContracted}. We have thus shown that unitary representations that contain non-trivial higher spin states are not possible for inherently non-linear flat space higher spin algebras, at least not under the assumptions we have been working with.



\subsection{Elimination of multi-graviton excitations ($\mathcal{W}_4^{(2-2)}$)}\label{se:4.4}

In this subsection we show that our no-go result also can eliminate multi-graviton excitations, if we demand flat space and unitarity. We focus on a specific example.

Namely, we Galilei-contract two copies of the $\mathcal{W}_4^{(2-2)}$-algebra, whose non-vanishing commutators read
\begin{subequations}\label{eq:W22QuantumAlgebra}
\begin{align}
[\mathcal{L}_n,\,\mathcal{L}_m]=&(n-m)\mathcal{L}_{n+m}+\frac{c}{12}n(n^2-1)\,\delta_{n+m,\,0}\\
[\mathcal{L}_n,\,T^{a}_m]=&(n-m)T^{a}_{n+m}\\
[\mathcal{L}_n,\,S^{a}_m]=&-mS^{a}_{n+m} \displaybreak[1] \\
[S^{a}_n,\,S^{b}_m]=&(a-b)S^{a+b}_{n+m}+\kappa(1-3a^2)n\,\delta_{a+b,\,0}\delta_{n+m,\,0}\label{eq:W22QuantumAlgebraSS}\\
[S^{a}_n,\,T^{b}_m]=&f(a,b)T^{a+b}_{n+m}\\
[T^{\pm}_n,\,T^{\pm}_m]=&\frac{3\kappa+2}{8 \kappa(\kappa+2)}(n-m):SS:^{\pm|\pm}_{n+m} \displaybreak[1] \\
[T^{0}_n,\,T^{0}_m]=&\frac{1}{4}(n-m)\mathcal{L}_{n+m}-\frac{\kappa-2}{8\kappa (\kappa+2)}(n-m):SS:_{n+m}^{0|0}\nonumber\\
&+\frac{1}{4 (\kappa+2)}(n-m):SS:_{n+m}^{\{+|-\}}-\frac{(2\kappa-1) (3\kappa+2)}{24 (\kappa+2)}n(n^2-1)\,\delta_{n+m,\,0}\label{eq:W22QuantumAlgebraNastyOne} \displaybreak[1] \\
[T^{\pm}_n,\,T^{0}_m]=&q(n,m)S^{\pm}_{n+m}\pm\left(\frac{\kappa-2}{8 \kappa (\kappa+2)}n-\frac{1}{4\kappa}m\right):SS:^{0|\pm}_{n+m}\pm\frac{1}{8 \kappa}:\Omega SS:^{0|\pm}_{n+m}\nonumber\\
&\pm\left(\frac{1}{4 \kappa}n-\frac{\kappa-2}{8\kappa (\kappa+2)}m\right):SS:^{\pm|0}_{n+m}\mp\frac{1}{8\kappa}:\Omega SS:^{\pm|0}_{n+m}\nonumber\\
&+\frac{1}{4 \kappa}\{:LS:\}^\pm_{n+m}+\frac{1}{6 \kappa(\kappa+2)}\left(:SSS:^{\{\pm|\pm|\mp\}}_{n+m}-:SSS:^{\{0|0|\pm\}}_{n+m}\right)\label{eq:W22QuantumAlgebraNastyOne1} \displaybreak[1] \\
[T^{-}_n,\,T^{+}_m]=&\frac{1}{2}(n-m)\mathcal{L}_{n+m}+2q(n,m)S^{0}_{n+m}\nonumber\\
&-\frac{(2\kappa-1) (3\kappa+2)}{12 (\kappa+2)}n(n^2-1)\delta_{n+m,0}-\frac{1}{\kappa+2}(n-m):SS:_{n+m}^{0|0}\nonumber\\
&+\left(\frac{3\kappa+2}{8\kappa(\kappa+2)}n-\frac{\kappa-2}{4\kappa(\kappa+2)}m\right):SS:_{n+m}^{+|-}-\frac{1}{8 \kappa}:\Omega SS:_{n+m}^{+|-}\nonumber\\
&+\left(\frac{\kappa-2}{4\kappa(\kappa+2)}n-\frac{3\kappa+2}{8\kappa (\kappa+2)}m\right):SS:_{n+m}^{-|+}+\frac{1}{8 \kappa}:\Omega SS:_{n+m}^{-|+}\nonumber\\
&+\frac{1}{2\kappa}\{:LS:\}_{n+m}^0+\frac{1}{6\kappa(\kappa+2)}:SSS:_{n+m}^{\{-|+|0\}}-\frac{1}{\kappa(\kappa+2)}:SSS:_{n+m}^{0|0|0}\label{eq:W22QuantumAlgebraNastyOne2}
\end{align}
\end{subequations}
with
\begin{align}
q(n,m)=&\frac{3\kappa^2-3\kappa-8}{12 \kappa (\kappa+2)}-\frac{12 \kappa^2+9 \kappa+22}{48 \kappa (\kappa+2)}(n^2+m^2) \nonumber\\&
+\frac{3 \kappa^2-3 \kappa-14}{12 \kappa (\kappa+2)}nm-\frac{7 \kappa+18}{16 \kappa (\kappa+2)}(n+m)\\
\kappa=&\frac{1}{24} \left(7-c-\sqrt{c^2-110 c+145}\right)
\end{align} 
and
\begin{align}
f(a,b)=&(a-b)(1+a(1-a+2ab))\\
AB^{a|b}_{n}=&\sum_{p\in\mathbb{Z}}A^a_{n-p}B^b_p\\
\Omega AB^{a|b}_{n}=&\sum_{p\in\mathbb{Z}}pA^a_{n-p}B^b_p\\
ABC^{a|b|c}_n=&\sum_{p,q\in\mathbb{Z}}A^a_{n-p-q}B^b_pC^c_q\\
AB^{\{a|b\}}_{n}=&AB^{a|b}_{n}+AB^{b|a}_{n}\\
ABC^{\{a|b|c\}}_n=&ABC^{a|b|c}_n+ABC^{c|a|b}_n+ABC^{b|c|a}_n+ABC^{a|c|b}_n+ABC^{c|b|a}_n+ABC^{b|a|c}_n\\
\{:LS:\}^{a}_n=&:LS:^{a}_n+:SL:^{a}_n
\end{align}
We now take two copies of \eqref{eq:W22QuantumAlgebra} with generators $\mathcal{L}_n,T^a_n,S^a_n$ and $\bar{\mathcal{L}}_n,\bar{T}^a_n,\bar{S}^a_n$ respectively and define the usual linear combinations.
	\begin{subequations}
		\begin{align}\label{eq:NRLinComb22}
			L_n:=&\mathcal{L}_n+\bar{\mathcal{L}}_{n}&&\quad M_n:=-\epsilon\left(\mathcal{L}_n-\bar{\mathcal{L}}_{n}\right)\\
			O^a_n:=&S^a_n+\bar{S}^a_{n}&&\quad P^a_n:=-\epsilon\left(S^a_n-\bar{S}^a_{n}\right)\\
			U^{a}:=&T^{a}_n+\bar{T}^{a}_{n}&&\quad V^{a}:=-\epsilon\left(T^{a}_n-\bar{T}^{a}_{n}\right)
		\end{align}
	\end{subequations}
The Galilean contraction $\epsilon\to 0$ yields the following algebra.
	\begin{subequations}\label{eq:FlatW32QuantumRescaled1}
		\begin{align}
			[L_n,\,L_m]=&(n-m)L_{n+m}+\frac{c_L}{12}n(n^2-1)\,\delta_{n+m,\,0}\\
			[L_n,\,M_m]=&(n-m)M_{n+m}+\frac{c_M}{12}n(n^2-1)\,\delta_{n+m,\,0} \displaybreak[1] \\
			[L_n,\,O^a_m]=&-mO^a_{n+m}\\	
			[L_n,\,P^a_m]=&-mP^a_{n+m}\\
			[L_n,\,U^a_m]=&(n-m)U^a_{n+m}\\
			[L_n,\,V^a_m]=&(n-m)V^a_{n+m}\\
			[M_n,\,O^a_m]=&-mP^a_{n+m}\\
			[M_n,\,U^a_m]=&(n-m)V^a_{n+m} \displaybreak[1] \\
			[O^a_n,\,O^b_m]=&(a-b)O^{a+b}_{n+m}+\frac{62-c_L}{12}(1-3a^2)n\,\delta_{a+b,\,0}\delta_{n+m,\,0}\\
			[O^a_n,\,P^b_m]=&(a-b)P^{a+b}_{n+m}-\frac{c_M}{12}(1-3a^2)\,\delta_{a+b,\,0}\delta_{n+m,\,0} \displaybreak[1] \\
			[O^a_n,\,U^b_m]=&f(a,b)U^{a+b}_{n+m}\\
			[O^a_n,\,V^b_m]=&f(a,b)V^{a+b}_{n+m}\\
			[P^a_n,\,U^b_m]=&f(a,b)V^{a+b}_{n+m} \displaybreak[1] \\
			[U^\pm_n,\,U^\pm_m]=&\frac{9(c_L-94)}{2c_M^2}(n-m):PP:^{\pm|\pm}_{n+m}+\frac{9}{2c_M}(n-m)\{:OP:\}^{\pm|\pm}_{n+m} \displaybreak[1] \\
			[U^0_n,\,U^0_m]=&\frac{1}{4}(n-m)L_{n+m}+\frac{474-3c_L}{2c_M^2}(n-m):PP:^{0|0}_{n+m}+\frac{3}{2c_M}\{:OP:\}^{0|0}_{n+m}\nonumber\\
					&+\frac{3(c_L-110)}{c_M^2}(n-m):PP:^{\{+|-\}}_{n+m}-\frac{3}{c_M}\{:OP:\}^{\{+|-\}}_{n+m}\nonumber\\
					&+\frac{c_L-18}{48}n(n^2-1)\,\delta_{n+m,\,0}  \displaybreak[1] \\
			[U^\pm_n,U^0_m]=&-\frac{1}{4}g(n,m)O^\pm_{n+m}+p(n,m)P^\pm_{n+m}\pm n\left(\frac{3(c_L-158)}{2c_M^2}:PP:^{0|\pm}_{n+m}-\frac{3}{2c_M}\{:OP:\}^{0|\pm}_{n+m}\right)\nonumber\\
					&\mp m\left(\frac{3(c_L-62)}{c_M^2}:PP:^{0|\pm}_{n+m}-\frac{3}{c_M}\{:OP:\}^{0|\pm}_{n+m}\right)\nonumber\\
					&\pm\left(\frac{3(c_L-62)}{2c_M^2}:\Omega PP:^{[0|\pm]}_{n+m}-\frac{3}{2c_M}\{:\Omega OP:\}^{[0|\pm]}_{n+m}\right)\nonumber\\
					&\pm n\left(\frac{3(c_L-62)}{c_M^2}:PP:^{\pm|0}_{n+m}-\frac{3}{c_M}\{:OP:\}^{\pm|0}_{n+m}\right)\nonumber\\
					&\mp m\left(\frac{3(c_L-158)}{2c_M^2}:PP:^{\pm|0}_{n+m}-\frac{3}{2c_M}\{:OP:\}^{\pm|0}_{n+m}\right)\nonumber\\
					&+\frac{3(c_L-62)}{2c_M^2}\{:MP:\}^\pm_{n+m}-\frac{3}{2c_M}\left(\{:MO:\}^\pm_{n+m}+\{:LP:\}^\pm_{n+m}\right)\nonumber\\
					&+\frac{24}{c_M^2}\left(\{:PPO:\}^{\{\pm|\pm|\mp\}}_{n+m}-\{:PPO:\}^{\{0|0|\pm\}}_{n+m}\right)\nonumber\\
					&-\frac{48(c_L-86)}{c_M^3}\left(:PPP:^{\{\pm|\pm|\mp\}}-:PPP:^{\{0|0|\pm\}}_{n+m}\right) \displaybreak[0] \\
			[U^-_n,U^+_m]=&\frac{1}{2}(n-m)L_{n+m}-\frac{1}{2}g(n,m)O^0_{n+m}+2p(n,m)P^0_{n+m}+\frac{c_L-18}{24}n(n^2-1)\delta_{n+m,0}\nonumber\\
					&+(n-m)\left(-\frac{12(c_L-110)}{c_M^2}:PP:^{0|0}_{n+m}+\frac{12}{c_M}\{:OP:\}^{0|0}_{n+m}\right)\nonumber\\
					& +n\left(\frac{9(c_L-94)}{2c_M^2}:PP:^{+|-}_{n+m}-\frac{9}{2c_M}\{:OP:\}^{+|-}_{n+m}\right)\nonumber\\
					&+m\left(\frac{(474-3c_L)}{c_M^2}:PP:^{+|-}_{n+m}+\frac{3}{c_M}\{:OP:\}^{+|-}_{n+m}\right)\nonumber\\
					&-\left(\frac{3(c_L-62)}{2c_M^2}:\Omega PP:^{[+|-]}_{n+m}-\frac{3}{2c_M}\{:\Omega OP:\}^{[+|-]}_{n+m}\right)\nonumber\\
					&-n\left(\frac{(474-3c_L)}{c_M^2}:PP:^{-|+}_{n+m}+\frac{3}{c_M}\{:OP:\}^{-|+}_{n+m}\right)\nonumber\\
					&-m\left(\frac{9(c_L-94)}{2c_M^2}:PP:^{-|+}_{n+m}-\frac{9}{2c_M}\{:OP:\}^{-|+}_{n+m}\right)\nonumber\\
					&+\frac{3(c_L-62)}{c_M^2}\{:MP:\}^0_{n+m}-\frac{3}{c_M}\left(\{:MO:\}^0_{n+m}+\{:LP:\}^0_{n+m}\right)\nonumber\\
					&+\frac{24}{c_M^2}\left(\{:PPO:\}^{\{-|+|0\}}_{n+m}-6\{:PPO:\}^{0|0|0}_{n+m}\right)\nonumber\\
					&-\frac{48(c_L-86)}{c_M^3}\left(:PPP:^{\{-|+|0\}}-6:PPP:^{0|0|0}_{n+m}\right)  \displaybreak[1] \\
			[U^\pm_n,V^\pm_m]=&\frac{9}{2c_M}(n-m):PP:^{\pm|\pm}_{n+m}\\
			[U^0_n,V^0_m]=&\frac{1}{4}(n-m)M_{n+m}+\frac{3}{2c_M}(n-m):PP:^{0|0}_{n+m}-\frac{3}{c_M}(n-m):PP:^{\{+|-\}}_{n+m}\nonumber\\
					&+\frac{c_M}{48}n(n^2-1)\delta_{n+m,\,0} \displaybreak[0] \\
			[U^\pm_n,V^0_m]=&-\frac{1}{4}g(n,m)P^\pm_{n+m}\mp\frac{3}{2c_M}n:PP:^{0|\pm}_{n+m}\pm\frac{3}{c_M}m:PP:^{0|\pm}_{n+m}\nonumber\\
					&\mp\frac{3}{2c_M}:\Omega PP:^{[0|\pm]}_{n+m}\mp\frac{3}{c_M}n:PP:^{\pm|0}_{n+m}\pm\frac{3}{2c_M}m:PP:^{\pm|0}_{n+m}\nonumber\\
					&-\frac{3}{2c_M}\{:MP:\}^\pm_{n+m}+\frac{24}{c_M^2}\left(:PPP:^{\{\pm|\pm|\mp\}}-:PPP:^{\{0|0|\pm\}}_{n+m}\right) \displaybreak[0] \\
			[U^-_n,V^+_m]=&\frac{1}{2}(n-m)M_{n+m}-\frac{1}{2}g(n,m)P^0_{n+m}+\frac{c_M}{24}n(n^2-1)\delta_{n+m,0}\nonumber\\
					&(n-m)\frac{12}{c_M}:PP:^{0|0}_{n+m}-\frac{9}{2c_M}n:PP:^{+|-}_{n+m}+\frac{3}{c_M}m:PP:^{+|-}_{n+m}\nonumber\\
					&-\frac{3}{2c_M}:\Omega PP:^{[+|-]}_{n+m}-\frac{3}{c_M}n:PP:^{-|+}_{n+m}+\frac{9}{2c_M}n:PP:^{-|+}_{n+m}\nonumber\\
					&-\frac{3}{c_M}\{:MP:\}^0_{n+m}+\frac{24}{c_M^2}\left(:PPP:^{\{-|+|0\}}-6:PPP:^{0|0|0}_{n+m}\right)
		\end{align}
	\end{subequations}
with
	\begin{subequations}
		\begin{align}
			p(n,m)=&\frac{18}{c_M}(1+nm)-\frac{15}{c_M^2}(n^2+m^2)-\frac{21}{c_M}(n+m)\\
			\{:AB:\}^{a|b}_n=&:AB:^{a|b}_n+:BA:^{a|b}_n\\
			\{:\Omega AB:\}^{a|b}_n=&:\Omega AB:^{a|b}_n+:\Omega BA:^{a|b}_n\\
			\{:ABC:\}^{a|b|c}_n=&:ABC:^{a|b|c}_n+:ACB:^{a|b|c}_n+:BAC:^{a|b|c}_n\nonumber\\
						&+:BCA:^{a|b|c}_n+:CAB:^{a|b|c}_n+:CBA:^{a|b|c}_n\\
			:\Omega AB:^{[a|b]}_n=&:\Omega AB:^{a|b}_n-:\Omega AB:^{b|a}_n\,.
		\end{align}
	\end{subequations}

As in the cases treated before we need $c_M=0$ in order to have a chance at finding unitary representations. Thus we make the following rescaling
	\begin{subequations}
		\begin{align}
			U^a_n&\rightarrow\hat{U}^a_n=\left(c_M\right)^\frac{3}{2}U^a_n,\\
			V^a_n&\rightarrow\hat{V}^a_n=\sqrt{c_M}V^a_n,
		\end{align}
	\end{subequations}
and take the limit $c_M\rightarrow0$. This yields the following non-vanishing commutation relations.
	\begin{subequations}\label{eq:FlatW32QuantumCMRescaled}
		\begin{align}
			[L_n,\,L_m]=&(n-m)L_{n+m}+\frac{c_L}{12}n(n^2-1)\,\delta_{n+m,\,0}\\
			[L_n,\,M_m]=&(n-m)M_{n+m} \displaybreak[1] \\
			[L_n,\,O^a_m]=&-mO^a_{n+m}\\	
			[L_n,\,P^a_m]=&-mP^a_{n+m}\\
			[L_n,\,\hat{U}^a_m]=&(n-m)\hat{U}^a_{n+m}\\
			[L_n,\,\hat{V}^a_m]=&(n-m)\hat{V}^a_{n+m}\\
			[M_n,\,O^a_m]=&-mP^a_{n+m} \displaybreak[1] \\
			[O^a_n,\,O^b_m]=&(a-b)O^{a+b}_{n+m}+\frac{62-c_L}{12}(1-3a^2)n\,\delta_{a+b,\,0}\delta_{n+m,\,0} \label{eq:fshsg20}\\
			[O^a_n,\,P^b_m]=&(a-b)P^{a+b}_{n+m}\\
			[O^a_n,\,\hat{U}^b_m]=&f(a,b)\hat{U}^{a+b}_{n+m}\\
			[O^a_n,\,\hat{V}^b_m]=&f(a,b)\hat{V}^{a+b}_{n+m} \displaybreak[1] \\
			[\hat{U}^\pm_n,\,\hat{U}^0_m]=&-48(c_L-86)\left(:PPP:^{\{\pm|\pm|\mp\}}-:PPP:^{\{0|0|\pm\}}_{n+m}\right)\\
			[\hat{U}^-_n,\,\hat{U}^+_m]=&-48(c_L-86)\left(:PPP:^{\{-|+|0\}}-6:PPP:^{0|0|0}_{n+m}\right)\\
			[\hat{U}^\pm_n,\,\hat{V}^0_m]=&24\left(:PPP:^{\{\pm|\pm|\mp\}}-:PPP:^{\{0|0|\pm\}}_{n+m}\right)\\
			[\hat{U}^-_n,\,\hat{V}^+_m]=&24\left(:PPP:^{\{-|+|0\}}-6:PPP:^{0|0|0}_{n+m}\right)
		\end{align}
	\end{subequations}

Using the standard definitions of the vacuum and hermitian conjugation and looking at the commutator \eqref{eq:fshsg20} by analogy to section \ref{se:4.2} we obtain again similar bounds on the central charge $c_L$ consistent with unitarity given by
	\begin{equation}
		\frac{1}{2} \big(77-\sqrt{5185}\big)\approx2.497\leq c_L \leq 62\,.
\label{eq:whatever2}
	\end{equation}
As before, only a discrete set of values for $c_L$ is allowed from the quantization of the $\hat{\mathfrak{su}}(2)_k$ level $k$, namely $c_L=6n+2$ with $n=1,\,2,\,\dots,\,10$. Moreover, again the lower bound in \eqref{eq:whatever2} emerges from non-negativity of the bare central charge, $c_{\textrm{\tiny bare}}=c_L-3(62-c_L)/(74-c_L)\geq 0$.
In addition, we see from the contracted algebra \eqref{eq:FlatW32QuantumCMRescaled} that the only non-trivial vacuum descendants are $L_{-n}|0\rangle$ with $n>1$, $O_{-n}^a|0\rangle$ with $n>0$ or combinations thereof. All other vacuum descendants are null states, in particular the ones generated by the other spin-2 generators $U_n^{\pm,0}$ and $V_n^{\pm,0}$. 

Thus, at least for the present example unitarity in flat space eliminates multi-graviton states. The generalization of this statement to arbitrary ${\cal W}$-algebras that lead to multiple spin-2 states should work along the lines of our no-go result in section \ref{se:4.1}.

\section{Yes-go: Unitarity in flat space hs$(1)$}\label{se:5}

The previous examples, as well as the general no-go argument of section \ref{se:4.1}, make it clear that unitarity (under the assumptions we work with in this paper) is not compatible with having non-trivial higher spin states for flat space higher spin algebras that are inherently non-linear. The non-linear character of the algebras is however crucial for the argument. In this section, we will show that focusing on linear higher spin algebras can easily evade the no-go result of \ref{se:4.1}.

An easy way to obtain linear flat space higher spin algebras is by performing an \.In\"on\"u--Wigner contraction of two copies of a linear AdS higher spin algebra of $\mathcal{W}$-type. $\mathcal{W}$-algebras are typically non-linear, but a few examples in which the algebra is isomorphic to a linear one are known. In this section we shall focus on a particular example, namely the Pope--Romans--Shen $\mathcal{W}_\infty$ algebra \cite{Pope:1989ew,Pope:1989sr}. In particular, we shall show that an \.In\"on\"u--Wigner contraction of two copies of this algebra exists that leads to a linear flat space higher spin algebra, for which the no-go theorem of section \ref{se:4.1} is evaded.

\subsection{The $\mathcal{W}_\infty$ algebra}

The $\mathcal{W}_\infty$ algebra is a linear, centrally extended, infinite-dimensional algebra, whose generators will be denoted by $V^i_m$, according to the conventions of \cite{Pope:1989sr}. In this notation, the index $m$ ranges over all integers, while the index $i$ takes on values over all natural numbers ($i=0,1,2,\cdots$). The commutation relations of the $\mathcal{W}_\infty$ algebra are explicitly given by
\begin{equation} \label{commutator}
\left[V^i_m, V^j_n \right] = \sum_{r=0}^{\left \lfloor{\frac{i+j}{2}} \right \rfloor} \, g^{ij}_{2r}(m,n)\, V^{i+j-2r}_{m+n} +  \, c^i(m) \, \delta^{ij}\, \delta_{m+n,0} \,.
\end{equation}
The sum extends up to the integer part of $(i+j)/2$, meaning that it ends with a term proportional to $V^1_{m+n}$ or $V^0_{m+n}$, depending on whether $i+j$ is odd or even. The central terms in the above algebra are determined by a single central charge $c$ in the following way:
\begin{equation} \label{centralterms}
c^i(m) = (m-i-1) (m-i)\cdots(m+i) (m+i+1)\, c^i \,, \qquad c^i = \frac{2^{2i-3} i! (i+2)!}{(2i+1)!!(2i+3)!!}\, c \,.
\end{equation}
The structure constants are given by
\begin{equation} \label{structconst}
g^{ij}_{2r}(m,n) = \frac{1}{2(2r+1)!} \, \phi^{ij}_{2r} \, N^{ij}_{2r}(m,n) \,,
\end{equation}
where $\phi^{ij}_{2r}$ is given in terms of a hypergeometric series
\begin{equation} \label{defphi}
\phi^{ij}_{2r} = \pFq{4}{3}{-\frac12,,,\frac32,,,-r-\frac12,,,-r}{-i-\frac12,,,-j-\frac12,,,,i+j-2r+\frac52}{1}\,,
\end{equation}
and $N^{ij}_{2r}(m,n)$ is given by
\begin{align} \label{defN}
N^{ij}_{2r}(m,n) & = \sum_{k=0}^{2r+1} (-1)^k \binom{2r+1}{k} \pocha{2i+2-2r}{k}\pochd{2j+2-k}{2r+1-k} \pochd{i+1+m}{2r+1-k}\nonumber \\ & \qquad \qquad  \pochd{j+1+n}{k} \,,
\end{align}
where we used the ascending and descending Pochhammer symbols $\pocha{a}{n}$, $\pochd{a}{n}$, defined in \eqref{defpoch}. There exists another way of writing $N^{ij}_{2r}(m,n)$ that is often useful:
\begin{align} \label{defN2}
N^{ij}_{2r}(m,n)& = \sum_{k=0}^{2r+1} (-1)^k \binom{2r+1}{k} \pochd{i+1+m}{2r+1-k} \pochd{i+1-m}{k} \pochd{j+1+n}{k}\nonumber \\ & \qquad \qquad \pochd{j+1-n}{2r+1-k} 
\end{align}

From the above expressions, it is easy to see that the generators $V^0_m$ form a subalgebra, whose commutation relation is given by
\begin{equation}
\left[V^0_m,V^0_n\right] = (m-n) V^0_{m+n} +  \frac{c}{12} (m^3-m) \delta_{m+n,0} \,.
\end{equation}
The generators $V^0_m \equiv L_m$ thus form a Virasoro subalgebra of $\mathcal{W}_\infty$. 

Another interesting subalgebra is the wedge algebra $\mathcal{W}_\infty^\wedge$, given by all generators $V^i_m$ that have $m$-indices for which the central terms vanish:
\begin{equation} \label{hs1}
\mathcal{W}_\infty^\wedge = \left\{V^i_m | -i-1\leq m \leq i+1 \right\} \,.
\end{equation}
This wedge algebra coincides with the higher spin algebra hs$(1)$ that corresponds to a special case of the hs$(\lambda)$ higher spin algebras. The latter are infinite-dimensional algebras that can be considered as suitable generalizations of sl$(N)$ to non-integer values of $N$ \cite{Bordemann:1989zi,Bergshoeff:1989ns,Vasiliev:1989re}. Similar to the hs$(1)$ case \eqref{hs1} above, hs$(\lambda)$ has generators $V^i_m$, with $|m| \leq i+1$, and $i=0,1,2,\cdots$. With respect to the sl$(2)$ subalgebra, formed by $V^0_{0}$ and $V^0_{\pm1}$, the generators $V^i_m$ transform in a representation of spin $i+1$.

The $\mathcal{W}_\infty$ algebra can be seen as a Virasoro-like extension of the hs$(1)$ algebra. In principle, one could start in three dimensions from a Chern-Simons theory based on the $\mathrm{hs}(1) \oplus \mathrm{hs}(1)$ algebra, generalizing the $\mathrm{sl}(N) \oplus \mathrm{sl}(N)$ case studied in \cite{Campoleoni:2010zq,Campoleoni:2011hg}. This theory describes an infinite number of higher spin fields in $\mathrm{AdS}_3$, of all integer spins higher than or equal to two. Imposing suitable boundary conditions, this Chern-Simons theory has an asymptotic symmetry algebra that corresponds to a Virasoro-like extension of the $\mathrm{hs}(1) \oplus \mathrm{hs}(1)$ algebra. An algebra consisting of two copies of the $\mathcal{W}_\infty$ algebra is thus a natural candidate for this asymptotic symmetry algebra. This is indeed the case, as was argued in \cite{Gaberdiel:2011wb}.

\subsection{Linear flat space higher spin algebra and unitarity}\label{se:fshscg}

Starting from two copies of the $\mathcal{W}_\infty$ algebra, with generators $V^i_m$, $\bar{V}^i_m$ and central charges $c$, $\bar{c}$, a flat space version of this algebra can be obtained via \.In\"on\"u--Wigner contraction. We perform an `ultra-relativistic' contraction, obtained by defining new generators $\mathcal{V}^i_m$, $\mathcal{W}^i_m$ using a dimensionful contraction parameter $\epsilon$ as follows:
\begin{equation}
\cV^i_m = V^i_m - \bar{V}^i_{-m} \,, \qquad \qquad
\cW^i_m = \epsilon \left( V^i_m + \bar{V}^i_{-m}\right) \,.
\label{eq:nolabel}
\end{equation}
Similarly, we define new central charges $c_\mathcal{V}$, $c_\mathcal{W}$:
\begin{equation}
c_{\mathcal{V}} = c - \bar{c} \,, \qquad \qquad
c_{\mathcal{W}} = \epsilon \left(c + \bar{c} \right) \,.
\end{equation}
The contraction is then performed by calculating commutators of the generators $\mathcal{V}^i_m$, $\mathcal{W}^i_m$ and taking the limit $\epsilon \rightarrow 0$. The resulting algebra is given by
\begin{align} \label{ihs1}
\left[\mathcal{V}^i_m,\mathcal{V}^j_n \right] & = \sum_{r=0}^{\left \lfloor{\frac{i+j}{2}} \right \rfloor} \, g^{ij}_{2r}(m,n)\, \mathcal{V}^{i+j-2r}_{m+n} +  \, c^i_{\mathcal{V}}(m) \, \delta^{ij}\, \delta_{m+n,0}  \nonumber \\
\left[\mathcal{V}^i_m,\mathcal{W}^j_n \right] & = \sum_{r=0}^{\left \lfloor{\frac{i+j}{2}} \right \rfloor} \, g^{ij}_{2r}(m,n)\, \mathcal{W}^{i+j-2r}_{m+n} +  \, c^i_{\mathcal{W}}(m) \, \delta^{ij}\, \delta_{m+n,0}  \nonumber \\
\left[\mathcal{W}^i_m,\mathcal{W}^j_n \right] & = 0 \,,
\end{align}
where the central charge terms $c^i_{\mathcal{V}}(m)$, $c^i_{\mathcal{W}}(m)$ are defined in a way analogous to \eqref{centralterms}.
In order to obtain this result, we used the following properties of the $\mathcal{W}_\infty$ structure constants and central charge terms:
\begin{equation}
g^{ij}_{2r}(-m,-n) = -g^{ij}_{2r}(m,n)\,, \qquad \qquad c^i(-m) = - c^i(m) \,.
\end{equation} 
These properties can be easily inferred from the explicit expressions \eqref{centralterms},  \eqref{structconst} and \eqref{defN2}.

The resulting algebra is linear and the generators $\mathcal{V}^0_m$, $\mathcal{W}^0_m$ form a BMS$_3$ subalgebra, with central charges $c_{\mathcal{V}}$ and $c_{\mathcal{W}}$, showing that this is indeed a flat space higher spin algebra. In order to discuss unitarity, we can look at the highest-weight representation, defined by the vacuum annihilation conditions:
\begin{align}
\mathcal{V}^i_m |0> = 0  \qquad \mathrm{and} \qquad \mathcal{W}^i_m |0> = 0 \,, \qquad m \geq -i-1 \,,
\end{align}
and the following definition of the adjoint:
\begin{equation}
\left(\mathcal{V}^{i}_m\right)^\dag = \mathcal{V}^i_{-m} \,, \qquad \qquad \left(\mathcal{W}^{i}_m\right)^\dag = \mathcal{W}^i_{-m}
\end{equation}
Since $c_{\mathcal{W}}$ plays the role of the $c_M$ central charge in the BMS$_3$ subalgebra of \eqref{ihs1}, unitarity can only be achieved when $c_{\mathcal{W}} \rightarrow 0$. This limit is now however regular and one finds that states created by $\mathcal{W}^i_m$ generators are null and can be modded out. For $c_{\mathcal{V}} \neq 0$, the only non-trivial states left are the ones created by $\mathcal{V}^i_m$. One is therefore left with states that are descendants of a single copy of a $\mathcal{W}_\infty$ algebra, that can however form a unitary representation of the flat space higher spin algebra \eqref{ihs1}. 


\subsection{Comments and generalizations}

The results of section \ref{se:fshscg} show that it is possible to have an asymptotic symmetry algebra in flat space higher spin gravity that allows unitary representations. An interesting aspect of our result is that this flat space higher spin theory has to be chiral, in the sense that there is a single copy of the higher spin generalization of the Virasoro algebra. Thus, a theory realizing this asymptotic symmetry algebra would be a higher spin analogue of flat space chiral gravity \cite{Bagchi:2012yk}. 

The example above is very special, but not unique. There is (at least) one other example that leads to the same conclusions. All one has to do is to replace the $\mathcal{W}_\infty$ algebra by the $\mathcal{W}_{1+\infty}$ algebra, which has an additional $\hat{\mathfrak{u}}(1)$ current algebra as compared to the $\mathcal{W}_\infty$ algebra \cite{Pope:1990kc}. The algebra is again linear and thereby circumvents our no-go result of section \ref{se:4.1}. Explicit results for the structure constants and central charge can be found e.g.~in Eqs.~(23)-(24) of \cite{Shen:1992dd} (whose conventions are compatible with ours).

We elaborate now further on the possibility of linearizing generic $W$-algebras by reconsidering the specific example of $W_3$. This algebra can be linearized by introducing infinitely many new generators (of higher spin), starting with $\Lambda_n = \sum_p \,\colon\!L_{n-p}L_p\colon$ and similarly for other non-linear terms that will show up in the commutators. However, the original \.In\"on\"u--Wigner contraction \eqref{eq:fshsg12} is then implemented non-linearly in terms of these new generators and leads to the same problem with poles in $1/c_M$ as in our general no-go result.
One could investigate alternative contractions. We have found that the straightforward linear combination of generators analogously to \eqref{eq:nolabel} does not lead to a consistent contraction here. We leave it as an open question whether or not alternative contractions exist. We expect that similar considerations hold for generic $W$-algebras, so that circumventions of the no-go result are exceptional cases, like the ones discussed in the present section.

Finally, we note that it is possible to linearize an extension of the $W_3$ algebra that includes a spin-1 current in a way that is analogous to the $W_\infty$ case \cite{Krivonos:1994qj}, i.e., by performing a non-linear change of basis. However, the results of Sorkin and Krivonos show that in this case the spin-3 field becomes a null field. Therefore, this example cannot be used neither to circumvent our no-go result.

\section{Conclusions}\label{se:6}

We showed by examples and by a general no-go result that for standard definitions of the vacuum and adjoint operators flat space higher spin gravity is incompatible with unitarity. Phrased differently, imposing unitarity leads to a further contraction of the asymptotic symmetry algebra that decouples all higher-spin states from the physical spectrum. Since every no-go result is only as good as its premises, we mention here again that a crucial input in the no-go result was the assumption of non-linearity in the asymptotic symmetry algebra.

Indeed, for higher-spin theories that allow a redefinition of generators in such a way that the asymptotic symmetry algebra linearizes we found that the no-go result can be circumvented. Two explicit examples of asymptotic symmetry algebras with unitary representations that could arise in flat space higher spin gravity were discussed in section \ref{se:5}.

We conclude with some outstanding questions and comments that point to possible future directions of research. While we have shown in section \ref{se:5} that flat space chiral higher spin gravity is unitary, if it exists, we have provided no evidence for its existence other than constructing its asymptotic symmetry algebra. However, as opposed to the spin-2 case where an explicit candidate for flat space chiral gravity is known on the gravity side --- namely conformal Chern--Simons gravity \cite{Bagchi:2012yk} --- we have no proposal for a higher spin version of that theory. Thus, an important open question is whether or not flat space chiral higher spin gravity exists. Technically, the problem is that the most natural constructions of flat space higher spin gravity always lead to asymptotic symmetry algebras where the wrong central charge is non-zero (namely the one which spoils unitarity). It is not clear to us what could be a reasonable candidate for flat space chiral higher spin gravity, but presumably it is a suitable generalization of conformal Chern--Simons gravity to a Vasiliev type theory.

An interesting reinterpretation of our results could provide a path towards unitarity even for truncated theories of higher spin gravity, i.e., theories that are not of Vasiliev type but have a finite tower of higher spin states. Namely, naively also the hs$(1)\oplus$\;hs$(1)$ theory leads to a non-linear asymptotic symmetry algebra, but there exists a non-linear
redefinition of the generators that leads to a linear algebra. The vacuum is then defined by highest weight conditions with respect to these new generators and not with respect to the `original' ones. This could be taken as an indication that also in the case of non-linear higher spin algebras (or even in the spin-2 case) a more suitable definition of the flat space vacuum might exist that is compatible with unitarity. However, currently we have no suggestion how this vacuum should be defined.

\acknowledgments

We are grateful to Hamid Afshar, Arjun Bagchi, Thomas Creutzig, Stephane Detournay, Reza Fareghbal, Michael Gary, Yasuaki Hikida, Eric Perlmutter, Stefan Prohazka, Radoslav Rashkov, Soo-Jong Rey, Peter R\o{}nne, Friedrich Sch\"oller, Joan Simon for collaboration on related topics. 
Additionally, we thank the participants of the workshop ``Higher-Spin and Higher-Curvature Gravity'' in S\~ao Paulo in November 2013 for many fruitful discussions.

DG, MR and JR were supported by the START project Y~435-N16 of the Austrian Science Fund (FWF) and the FWF projects I~952-N16 and I~1030-N27. 
DG's extended visit to S\~ao Paulo was supported by FAPESP.
DG additionally thanks the Asia Pacific Center for Theoretical Physics (APCTP), the Centro de Ciencias de Benasque, ABC do Federal University, ICTP SAIFR, the University of Buenos Aires and IAFE for hospitality during the preparation of this work. 

\appendix

\section{Pochhammer symbols}

Unless otherwise stated, $n$ denotes a general non-negative integer.
The ascending and descending Pochhammer symbols are denoted by $\pocha{a}{n}$, $\pochd{a}{n}$, respectively, and are defined by
\begin{subequations}
 \label{defpoch}
\begin{align} 
\pocha{a}{n} &\equiv a (a+1) \cdots (a + n -1) \,,  \\
\pochd{a}{n} &\equiv a (a-1) \cdots (a- n + 1) \,,
\end{align}
\end{subequations}
with $\pocha{a}{0} \equiv \pochd{a}{0} \equiv 1$. Note that $\pocha{a}{n}$ and $\pochd{a}{n}$ are both polynomials in $a$ of degree $n$. It is straightforward to see that
\begin{equation} \label{proppoch}
\pochd{a}{n} = \pocha{a-n+1}{n} = (-1)^n \pocha{-a}{n} \,.
\end{equation}
Even though in this text $n$ is a non-negative integer, it is possible to define Pochhammer symbols for arbitrary $n$, i.e., to analytically continue in $n$, by first expressing the definition of the Pochhammer symbols in terms of factorials and by expressing the ensuing expressions in terms of $\Gamma$-functions. Finally, note that
\begin{equation} \label{pochdzero}
\pochd{a}{n} = 0 \, \qquad \mathrm{for}\, \, n > a \quad (n, a \  \mathrm{non}\mbox{-}\mathrm{negative}\ \mathrm{integers}) \,.
\end{equation}


\begin{thebibliography}{10}

\bibitem{Bekaert:2010hw}
X.~Bekaert, N.~Boulanger, and P.~Sundell, ``{How higher-spin gravity surpasses
  the spin two barrier: no-go theorems versus yes-go examples},'' {\em
  Rev.Mod.Phys.} {\bf 84} (2012) 987--1009,
\href{http://www.arXiv.org/abs/1007.0435}{{\tt 1007.0435}}.

\bibitem{Vasiliev:1990en}
M.~A. Vasiliev, ``{Consistent equation for interacting gauge fields of all
  spins in (3+1)-dimensions},'' {\em Phys.Lett.} {\bf B243} (1990) 378--382.

\bibitem{Bekaert:2005vh}
X.~Bekaert, S.~Cnockaert, C.~Iazeolla, and M.~Vasiliev, ``{Nonlinear higher
  spin theories in various dimensions},''
  \href{http://www.arXiv.org/abs/hep-th/0503128}{{\tt hep-th/0503128}}.

\bibitem{Didenko:2014dwa}
V.~Didenko and E.~Skvortsov, ``{Elements of Vasiliev theory},''
\href{http://www.arXiv.org/abs/1401.2975}{{\tt 1401.2975}}.

\bibitem{Afshar:2013vka}
H.~Afshar, A.~Bagchi, R.~Fareghbal, D.~Grumiller, and J.~Rosseel, ``{Higher
  spin theory in 3-dimensional flat space},'' {\em Phys.Rev.Lett.} {\bf 111}
  (2013) 121603,
\href{http://www.arXiv.org/abs/1307.4768}{{\tt 1307.4768}}.

\bibitem{Gonzalez:2013oaa}
H.~A. Gonzalez, J.~Matulich, M.~Pino, and R.~Troncoso, ``{Asymptotically flat
  spacetimes in three-dimensional higher spin gravity},'' {\em JHEP} {\bf 1309}
  (2013) 016,
\href{http://www.arXiv.org/abs/1307.5651}{{\tt 1307.5651}}.

\bibitem{Achucarro:1987vz}
A.~Achucarro and P.~K. Townsend, ``A {C}hern-{S}imons action for
  three-dimensional {A}nti-de {S}itter supergravity theories,'' {\em Phys.
  Lett.} {\bf B180} (1986)
89.

\bibitem{Witten:1988hc}
E.~Witten, ``(2+1)-dimensional gravity as an exactly soluble system,'' {\em
  Nucl. Phys.} {\bf B311} (1988)
46.

\bibitem{Campoleoni:2010zq}
A.~Campoleoni, S.~Fredenhagen, S.~Pfenninger, and S.~Theisen, ``{Asymptotic
  symmetries of three-dimensional gravity coupled to higher-spin fields},''
  {\em JHEP} {\bf 1011} (2010) 007,
  \href{http://www.arXiv.org/abs/1008.4744}{{\tt 1008.4744}}.

\bibitem{Campoleoni:2011hg}
A.~Campoleoni, S.~Fredenhagen, and S.~Pfenninger, ``{Asymptotic W-symmetries in
  three-dimensional higher-spin gauge theories},'' {\em JHEP} {\bf 1109} (2011)
  113,
\href{http://www.arXiv.org/abs/1107.0290}{{\tt 1107.0290}}.

\bibitem{Pope:1989ew}
C.~Pope, L.~Romans, and X.~Shen, ``{The Complete Structure of W(Infinity)},''
  {\em Phys.Lett.} {\bf B236} (1990)
173.

\bibitem{Pope:1989sr}
C.~Pope, L.~Romans, and X.~Shen, ``{$W$(infinity) and the Racah-Wigner
  Algebra},'' {\em Nucl.Phys.} {\bf B339} (1990)
191--221.

\bibitem{Brown:1986nw}
J.~D. Brown and M.~Henneaux, ``{Central Charges in the Canonical Realization of
  Asymptotic Symmetries: An Example from Three-Dimensional Gravity},'' {\em
  Commun. Math. Phys.} {\bf 104} (1986)
207--226.

\bibitem{Barnich:2006av}
G.~Barnich and G.~Compere, ``{Classical central extension for asymptotic
  symmetries at null infinity in three spacetime dimensions},'' {\em
  Class.Quant.Grav.} {\bf 24} (2007) F15--F23,
\href{http://www.arXiv.org/abs/gr-qc/0610130}{{\tt gr-qc/0610130}}.

\bibitem{Bagchi:2009my}
A.~Bagchi and R.~Gopakumar, ``{Galilean Conformal Algebras and AdS/CFT},'' {\em
  JHEP} {\bf 0907} (2009) 037,
\href{http://www.arXiv.org/abs/0902.1385}{{\tt 0902.1385}}.

\bibitem{Bagchi:2009pe}
A.~Bagchi, R.~Gopakumar, I.~Mandal, and A.~Miwa, ``{GCA in 2d},'' {\em JHEP}
  {\bf 1008} (2010) 004,
\href{http://www.arXiv.org/abs/0912.1090}{{\tt 0912.1090}}.

\bibitem{Duval:2014uoa}
C.~Duval, G.~Gibbons, P.~Horvathy, and P.~Zhang, ``{Carroll versus Newton and
  Galilei: two dual non-Einsteinian concepts of time},''
\href{http://www.arXiv.org/abs/1402.0657}{{\tt 1402.0657}}.

\bibitem{Duval:2014uva}
C.~Duval, G.~Gibbons, and P.~Horvathy, ``{Conformal Carroll groups and BMS
  symmetry},''
\href{http://www.arXiv.org/abs/1402.5894}{{\tt 1402.5894}}.

\bibitem{Duval:2014lpa}
C.~Duval, G.~Gibbons, and P.~Horvathy, ``{Conformal Carroll groups},''
\href{http://www.arXiv.org/abs/1403.4213}{{\tt 1403.4213}}.

\bibitem{Bagchi:2010zz}
A.~Bagchi, ``{Correspondence between Asymptotically Flat Spacetimes and
  Nonrelativistic Conformal Field Theories},'' {\em Phys.Rev.Lett.} {\bf 105}
  (2010)
171601.

\bibitem{Bagchi:2012cy}
A.~Bagchi and R.~Fareghbal, ``{BMS/GCA Redux: Towards Flatspace Holography from
  Non-Relativistic Symmetries},'' {\em JHEP} {\bf 1210} (2012) 092,
\href{http://www.arXiv.org/abs/1203.5795}{{\tt 1203.5795}}.

\bibitem{Barnich:2010eb}
G.~Barnich and C.~Troessaert, ``{Aspects of the BMS/CFT correspondence},'' {\em
  JHEP} {\bf 1005} (2010) 062,
\href{http://www.arXiv.org/abs/1001.1541}{{\tt 1001.1541}}.

\bibitem{Barnich:2011mi}
G.~Barnich and C.~Troessaert, ``{BMS charge algebra},'' {\em JHEP} {\bf 1112}
  (2011) 105,
\href{http://www.arXiv.org/abs/1106.0213}{{\tt 1106.0213}}.

\bibitem{Barnich:2012aw}
G.~Barnich, A.~Gomberoff, and H.~A. Gonzalez, ``{The Flat limit of three
  dimensional asymptotically anti-de Sitter spacetimes},'' {\em Phys.Rev.} {\bf
  D86} (2012) 024020,
\href{http://www.arXiv.org/abs/1204.3288}{{\tt 1204.3288}}.

\bibitem{Cornalba:2002fi}
L.~Cornalba and M.~S. Costa, ``{A New cosmological scenario in string
  theory},'' {\em Phys.Rev.} {\bf D66} (2002) 066001,
\href{http://www.arXiv.org/abs/hep-th/0203031}{{\tt hep-th/0203031}}.

\bibitem{Simon:2002ma}
J.~Simon, ``{The Geometry of null rotation identifications},'' {\em JHEP} {\bf
  0206} (2002) 001,
\href{http://www.arXiv.org/abs/hep-th/0203201}{{\tt hep-th/0203201}}.

\bibitem{Cornalba:2003kd}
L.~Cornalba and M.~S. Costa, ``{Time dependent orbifolds and string
  cosmology},'' {\em Fortsch.Phys.} {\bf 52} (2004) 145--199,
\href{http://www.arXiv.org/abs/hep-th/0310099}{{\tt hep-th/0310099}}.

\bibitem{Bagchi:2012yk}
A.~Bagchi, S.~Detournay, and D.~Grumiller, ``{Flat-Space Chiral Gravity},''
  {\em Phys.Rev.Lett.} {\bf 109} (2012) 151301,
\href{http://www.arXiv.org/abs/1208.1658}{{\tt 1208.1658}}.

\bibitem{Witten:2007kt}
E.~Witten, ``{Three-Dimensional Gravity Revisited},''
\href{http://www.arXiv.org/abs/0706.3359}{{\tt 0706.3359}}.

\bibitem{Li:2008dq}
W.~Li, W.~Song, and A.~Strominger, ``{Chiral Gravity in Three Dimensions},''
  {\em JHEP} {\bf 04} (2008) 082,
\href{http://www.arXiv.org/abs/0801.4566}{{\tt 0801.4566}}.

\bibitem{Barnich:2012xq}
G.~Barnich, ``{Entropy of three-dimensional asymptotically flat cosmological
  solutions},'' {\em JHEP} {\bf 1210} (2012) 095,
\href{http://www.arXiv.org/abs/1208.4371}{{\tt 1208.4371}}.

\bibitem{Bagchi:2012xr}
A.~Bagchi, S.~Detournay, R.~Fareghbal, and J.~Simon, ``{Holography of 3d Flat
  Cosmological Horizons},'' {\em Phys. Rev. Lett.} {\bf 110} (2013) 141302,
\href{http://www.arXiv.org/abs/1208.4372}{{\tt 1208.4372}}.

\bibitem{Barnich:2012rz}
G.~Barnich, A.~Gomberoff, and H.~A. Gonzalez, ``{BMS$_3$ invariant two
  dimensional field theories as flat limit of Liouville},'' {\em Phys. Rev.}
  {\bf D87:124032,} (2013)
\href{http://www.arXiv.org/abs/1210.0731}{{\tt 1210.0731}}.

\bibitem{Barnich:2013yka}
G.~Barnich and H.~A. Gonzalez, ``{Dual dynamics of three dimensional
  asymptotically flat Einstein gravity at null infinity},'' {\em JHEP} {\bf
  1305} (2013) 016,
\href{http://www.arXiv.org/abs/1303.1075}{{\tt 1303.1075}}.

\bibitem{Bagchi:2013lma}
A.~Bagchi, S.~Detournay, D.~Grumiller, and J.~Simon, ``{Cosmic evolution from
  phase transition of 3-dimensional flat space},''
\href{http://www.arXiv.org/abs/1305.2919}{{\tt 1305.2919}}.

\bibitem{Afshar:2013bla}
H.~R. Afshar, ``{Flat/AdS boundary conditions in three dimensional conformal
  gravity},'' {\em JHEP} {\bf 1310} (2013) 027,
\href{http://www.arXiv.org/abs/1307.4855}{{\tt 1307.4855}}.

\bibitem{Schulgin:2013xya}
W.~Schulgin and J.~Troost, ``{Asymptotic symmetry groups and operator
  algebras},'' {\em JHEP} {\bf 1309} (2013) 135,
\href{http://www.arXiv.org/abs/1307.3423}{{\tt 1307.3423}}.

\bibitem{Strominger:2013lka}
A.~Strominger, ``{Asymptotic Symmetries of Yang-Mills Theory},''
\href{http://www.arXiv.org/abs/1308.0589}{{\tt 1308.0589}}.

\bibitem{He:2014laa}
T.~He, V.~Lysov, P.~Mitra, and A.~Strominger, ``{BMS supertranslations and
  Weinberg's soft graviton theorem},''
\href{http://www.arXiv.org/abs/1401.7026}{{\tt 1401.7026}}.

\bibitem{Barnich:2013axa}
G.~Barnich and C.~Troessaert, ``{Comments on holographic current algebras and
  asymptotically flat four dimensional spacetimes at null infinity},'' {\em
  JHEP} {\bf 1311} (2013) 003,
\href{http://www.arXiv.org/abs/1309.0794}{{\tt 1309.0794}}.

\bibitem{Banks:2014iha}
T.~Banks, ``{The Super BMS Algebra, Scattering and Holography},''
\href{http://www.arXiv.org/abs/1403.3420}{{\tt 1403.3420}}.

\bibitem{Barnich:2013sxa}
G.~Barnich and P.-H. Lambert, ``{Einstein-Yang-Mills theory : I. Asymptotic
  symmetries},''
\href{http://www.arXiv.org/abs/1310.2698}{{\tt 1310.2698}}.

\bibitem{Wald:1999wa}
R.~M. Wald and A.~Zoupas, ``{A General definition of 'conserved quantities' in
  general relativity and other theories of gravity},'' {\em Phys.Rev.} {\bf
  D61} (2000) 084027,
\href{http://www.arXiv.org/abs/gr-qc/9911095}{{\tt gr-qc/9911095}}.

\bibitem{Costa:2013vza}
R.~N.~C. Costa, ``{Aspects of the zero $\Lambda$ limit in the AdS/CFT
  correspondence},''
\href{http://www.arXiv.org/abs/1311.7339}{{\tt 1311.7339}}.

\bibitem{Fareghbal:2013ifa}
R.~Fareghbal and A.~Naseh, ``{Flat-Space Energy-Momentum Tensor from BMS/GCA
  Correspondence},''
\href{http://www.arXiv.org/abs/1312.2109}{{\tt 1312.2109}}.

\bibitem{Bagchi:2013qva}
A.~Bagchi and R.~Basu, ``{3D Flat Holography: Entropy and Logarithmic
  Corrections},''
\href{http://www.arXiv.org/abs/1312.5748}{{\tt 1312.5748}}.

\bibitem{Krishnan:2013wta}
C.~Krishnan, A.~Raju, and S.~Roy, ``{A Grassmann Path From AdS$_3$ to Flat
  Space},''
\href{http://www.arXiv.org/abs/1312.2941}{{\tt 1312.2941}}.

\bibitem{Krishnan:2013tza}
C.~Krishnan and S.~Roy, ``{Desingularization of the Milne Universe},''
\href{http://www.arXiv.org/abs/1311.7315}{{\tt 1311.7315}}.

\bibitem{Detournay:2014fva}
S.~Detournay, D.~Grumiller, F.~Scholler, and J.~Simon, ``{Variational principle
  and 1-point functions in 3-dimensional flat space Einstein gravity},''
\href{http://www.arXiv.org/abs/1402.3687}{{\tt 1402.3687}}.

\bibitem{Henneaux:2010xg}
M.~Henneaux and S.-J. Rey, ``{Nonlinear $W_{infinity}$ as Asymptotic Symmetry
  of Three-Dimensional Higher Spin Anti-de Sitter Gravity},'' {\em JHEP} {\bf
  1012} (2010) 007, \href{http://www.arXiv.org/abs/1008.4579}{{\tt 1008.4579}}.

\bibitem{Zamolodchikov:1985wn}
A.~Zamolodchikov, ``{Infinite Additional Symmetries in Two-Dimensional
  Conformal Quantum Field Theory},'' {\em Theor.Math.Phys.} {\bf 65} (1985)
1205--1213.

\bibitem{Mathieu:1988pm}
P.~Mathieu, ``{Extended Classical Conformal Algebras and the Second Hamiltonian
  Structure of Lax Equations},'' {\em Phys.Lett.} {\bf B208} (1988)
101.

\bibitem{Bakas:1989mx}
I.~Bakas, ``{Hamiltonian Reduction and Conformal Symmetries in
  Two-dimensions},'' {\em Phys.Lett.} {\bf B219} (1989)
283--290.

\bibitem{Polyakov:1989dm}
A.~M. Polyakov, ``{Gauge Transformations and Diffeomorphisms},'' {\em
  Int.J.Mod.Phys.} {\bf A5} (1990)
833.

\bibitem{Bershadsky:1990bg}
M.~Bershadsky, ``{Conformal field theories via Hamiltonian reduction},'' {\em
  Commun.Math.Phys.} {\bf 139} (1991)
71--82.

\bibitem{Afshar:2012nk}
H.~Afshar, M.~Gary, D.~Grumiller, R.~Rashkov, and M.~Riegler, ``{Non-AdS
  holography in 3-dimensional higher spin gravity - General recipe and
  example},'' {\em JHEP} {\bf 1211} (2012) 099,
\href{http://www.arXiv.org/abs/1209.2860}{{\tt 1209.2860}}.

\bibitem{diFrancesco}
P.~Di~Francesco, P.~Mathieu, and D.~Senechal, {\em Conformal Field Theory}.
\newblock Springer, 1997.

\bibitem{Bordemann:1989zi}
M.~Bordemann, J.~Hoppe, and P.~Schaller, ``Infinite dimensional matrix
  algebras,'' {\em Phys.Lett.} {\bf B232} (1989)
199.

\bibitem{Bergshoeff:1989ns}
E.~Bergshoeff, M.~Blencowe, and K.~Stelle, ``Area preserving diffeomorphisms
  and higher spin algebra,'' {\em Commun.Math.Phys.} {\bf 128} (1990) 213.

\bibitem{Vasiliev:1989re}
M.~A. Vasiliev, ``{Higher Spin Algebras and Quantization on the Sphere and
  Hyperboloid},'' {\em Int.J.Mod.Phys.} {\bf A6} (1991)
1115--1135.

\bibitem{Gaberdiel:2011wb}
M.~R. Gaberdiel and T.~Hartman, ``{Symmetries of Holographic Minimal Models},''
  {\em JHEP} {\bf 1105} (2011) 031,
\href{http://www.arXiv.org/abs/1101.2910}{{\tt 1101.2910}}.

\bibitem{Pope:1990kc}
C.~Pope, L.~Romans, and X.~Shen, ``{A New Higher Spin Algebra and the Lone Star
  Product},'' {\em Phys.Lett.} {\bf B242} (1990)
401--406.

\bibitem{Shen:1992dd}
X.~Shen, ``{W infinity and string theory},'' {\em Int.J.Mod.Phys.} {\bf A7}
  (1992) 6953--6994,
\href{http://www.arXiv.org/abs/hep-th/9202072}{{\tt hep-th/9202072}}.

\bibitem{Krivonos:1994qj}
S.~Krivonos and A.~S. Sorin, ``{Linearizing W algebras},'' {\em Phys.Lett.}
  {\bf B335} (1994) 45--50,
\href{http://www.arXiv.org/abs/hep-th/9406005}{{\tt hep-th/9406005}}.

\end{thebibliography}
 
\providecommand{\href}[2]{#2}\begingroup\raggedright\endgroup

\end{document}